\documentclass[useAMS,usenatbib]{mn2e}

\pdfoutput=1 
\usepackage{graphicx}
\usepackage{color}
\usepackage{caption}
\usepackage{paralist} 
\usepackage{pdfpages}
\usepackage{ amssymb }

\newcommand{\msr}{$\mathcal{M_*}-R_{\rm e}\ $}

\DeclareGraphicsExtensions{.pdf,.png,.gif,.jpg}

\title[GAMA: \msr relations]
	{Galaxy And Mass Assembly (GAMA):  mass-size relations of z$<$0.1 galaxies subdivided by S\'ersic index, colour and morphology}

\author[R. Lange et al.] 
			{Rebecca Lange$^{1}$, Simon P. Driver$^{1,2}$, Aaron S.G. Robotham$^{1}$, Lee S. Kelvin$^{3}$, \newauthor Alister W. Graham$^{4}$, Mehmet Alpaslan$^{5}$, Stephen K. Andrews$^{1}$,  Ivan K. Baldry$^{6}$, \newauthor Steven Bamford$^{7}$,  Joss Bland-Hawthorn$^{8}$, Sarah Brough$^{9}$, Michelle E. Cluver$^{10}$, \newauthor Christopher J. Conselice$^{7}$, Luke J.M. Davies$^{1}$, Boris Haeussler$^{11,12}$,
\newauthor Iraklis S. Konstantopoulos$^{9}$,  Jon Loveday$^{13}$, Amanda J. Moffett$^{1}$,
\newauthor 	Peder Norberg$^{14}$, Steven Phillipps$^{15}$, Edward N. Taylor$^{16}$, 
\newauthor \'Angel R. L\'opez-S\'anchez$^{9}$, Stephen M. Wilkins$^{13}$ \\     
$^1$International Centre for Radio Astronomy Research (ICRAR), 
University of Western Australia, M468, 35 Stirling Highway,\\ Crawley, WA 6009, Australia \\
$^2$Scottish Universities' Physics Alliance (SUPA), School of Physics and 
Astronomy, University of St. Andrews, North Haugh,\\ St. Andrews, KY16 9SS, UK\\
$^3$Institut f\"{u}r Astro- und Teilchenphysik, Universit\"{a}t Innsbruck, 
Technikerstra{\ss}e 25, 6020 Innsbruck, Austria\\
$^{4}$Centre for Astrophysics and Supercomputing, Swinburne University of Technology, Hawthorn, Victoria 3122, Australia	\\
$^5$NASA Ames Research Center, MS 232, Moffett Field, CA 94035, USA\\
$^6$Astrophysics Research Institute, Liverpool John Moores University,
IC2, Liverpool Science Park, 146 Brownlow Hill, Liverpool, L3 5RF, UK\\
$^7$School of Physics and Astronomy, University of Nottingham, Nottingham NG7 2RD, UK\\
$^8$Sydney Institute for Astronomy, School of Physics A28, 
University of Sydney, NSW 2088, Australia\\
$^9$Australian Astronomical Observatory, P.O.~Box 915, North Ryde, 
NSW 1670, Australia \\
$^{10}$Astrophysics, Cosmology and Gravity Centre, University of Cape Town, Private Bag X3, Rondebosch 7701, Republic of South Africa 	\\
$^{11}$Department of Physics, University of Oxford, Denys Wilkinson Building, Keble Road, Oxford, Oxon, OX1 3RH, UK \\
$^{12}$University of Hertfordshire, Hatfield, Hertfordshire, AL10 9AB, UK	\\
$^{13}$Astronomy Centre, Department of Physics and Astronomy, 
University of Sussex, Brighton, BN1 9QH, UK	\\
$^{14}$ICC, Durham University, Durham, County Durham DH1 3, UK\\
$^{15}$School of Physics, University of Bristol, Bristol BS8 1TL, UK\\
$^{16}$School of Physics, the University of Melbourne, VIC 3010, Australia\\ 
}
\date{Submitted August 2014}
\pagerange{\pageref{firstpage}--\pageref{lastpage}}\pubyear{2014}

\begin{document}
\label{firstpage}
\maketitle

\begin{abstract}
We use data from the Galaxy And Mass Assembly (GAMA) survey in the redshift range 0.01$<$z$<$0.1 (8399 galaxies in $g$ to $K_s$ bands) to derive the stellar mass -- half-light radius relations for various divisions of `early' and `late'-type samples. We find the choice of division between early and late (i.e., colour, shape, morphology) is not particularly critical, however, the adopted mass limits and sample selections (i.e., the careful rejection of outliers and use of robust fitting methods) are important. In particular we note that for samples extending to low stellar mass limits ($<10^{10}\mathcal{M_{\sun}}$) the S\'ersic index bimodality, evident for high mass systems, becomes less distinct and no-longer acts as a reliable separator of early- and late-type systems. The final set of stellar mass -- half-light radius relations are reported for a variety of galaxy population subsets in 10 bands ($ugrizZYJHK_s$) and are intended to provide a comprehensive low-z benchmark for the many ongoing high-z studies. 
Exploring the variation of the stellar mass -- half-light radius relations with wavelength we confirm earlier findings that galaxies appear more compact at longer wavelengths albeit at a smaller level than previously noted: at $10^{10}\mathcal{M_{\sun}}$ both spiral systems and ellipticals show a decrease in size of 13\% from $g$ to $K_s$ (which is near linear in log wavelength). 
Finally we note that the sizes used in this work are derived from 2D S\'ersic light profile fitting (using GALFIT3), i.e., elliptical semi-major half light radii, improving on earlier low-z benchmarks based on circular apertures.
\end{abstract}


 \section{Introduction}
 
Galaxies have long been known to exhibit a correlation between their
mass (or luminosity) and their size (or surface brightness).  For example, early
studies of spiral galaxies in nearby groups and clusters identified a strong
luminosity-surface brightness relation, such that more luminous systems 
also have higher surface brightness (see reviews by
\citealt{Ferguson1994} and \citealt{Graham2013}). Fundamentally, this 
reflects the mean scaling of angular momentum with halo mass, which, 
in self-similar halos, is such that the disk surface density increases monotonically 
as $\mathcal{M}_{\rm halo}^{1/3}$ \citep{Fall1980,Dalcanton1997,Mo1998,Obreschkow2014}. 
The close connection between size (or surface brightness) and angular momentum 
makes systematic size measurements in galaxy surveys an exquisite test of 
evolutionary models \citep{Fall1983,Romanowsky2012}.\\

Over the past few decades, a number of notable observations have refined 
the empirical luminosity- surface brightness relation for distinct galaxy types 
(\citealt{DeJong2000, Graham2003}),
environments (\citealt{Cross2001, Andreon2002, Driver2005,
 Cappellari2013}) and at specific redshifts (\citealt{Driver1999,
LaBarbera2003, Barden2005, Trujillo2004, Trujillo2006, Trujillo2007,
Trujillo2012}).  

More recently, a growing number of authors choose to focus on the stellar mass -- half-light radius relation 
(hereafter \msr relation, see e.g. \citealt{Trujillo2004,Shen2003}) instead of 
a luminosity-surface brightness relation. 
Albeit akin to one another, the former is arguably 
more meaningful as the luminosity-size relation depends on the observational wavelength 
band used and conversions are required to compare different data sets. 
However, detractors of the \msr relation may argue that
this incorporates errors in the estimation of the stellar mass and
that the inherent selection boundaries (mainly due to surface
brightness selection effects, \citealt{Disney1976,Disney1995,Driver1999}) are less obvious in the \msr
plane than in the luminosity-surface brightness plane (and often
neglected altogether). 
For example, due to our inability to detect lower surface brightness 
sources at increasing redshifts only the more compact and most massive systems remain detectable, see e.g.~\cite{Cameron2007} who show the impact of the selection boundaries
using Hubble Space Telescope (HST) Ultra Deep Field (UDF) data.
Without due consideration of 
potential biases this becomes an important issue as differences
between the high- and low-redshift \msr relations are readily
attributed to physical evolution in galaxies.\\

A further, and more recent, concern is that the measured
size of a galaxy also depends on the wavelength at which
the observation has been made. This has been known for some time (e.g., \citealt{Evans1994,
Cunow2001, LaBarbera2002}), but quantified more robustly for red and/ or
blue systems in \cite{LaBarbera2010}, \cite{Kelvin2012}, \cite{Haussler2013} and \cite{Vulcani2014}, who find a strong size-wavelength relation, such that galaxies are often
measured to be as little as half the size in the K-band when compared with the \textit{r}-band. 
This is as crucial as the problems with the completeness discussed above. If one wishes to
measure and compare the \msr relation from different datasets or
from different epochs, care must be taken to define the relation at the same rest-wavelength or to apply a size bandpass correction (see \citealt{Kelvin2012}, figure 22).  
The cause of the size-wavelength trend (discussed in \citealt{Kelvin2012, Vulcani2014}) is not entirely clear, but is argued to arise from a combination of:
\begin{itemize}
\item the dust distribution, which preferentially blocks the central regions of galaxies 
(see for example the predictions by \citealt{Mollenhoff2006, Pastrav2013a}); 
\item the inside-out growth of
galaxies (where young bright stellar populations are more widely distributed
than the old stellar populations, \citealt{LaBarbera2010}); 
\item the two-component nature of many of the brightest galaxies (which consists of an old
centrally concentrated bulge superimposed on a young more diffuse
disc, i.e., the bulge is more evident in the K-band while the disc is
more evident in the \textit{r}-band, see \citealt{Driver2007b, Driver2007a});
\item and, to a much lesser degree, any metallicity gradients which may
also exist \citep{LaBarbera2010}.\\
\end{itemize}

Over the past decade observations of the \msr relation, particularly of early-type massive systems, have been made across a broad range of epochs using the high resolution imaging of the Advanced Camera for Surveys (ACS), the Wide Field and Planetary Camera 2 (WFPC2) or the Wide Field Camera 3 (WFC3) onboard the HST. 
These measurements, initially only made for the most
massive ($>10^{11}\mathcal{M}_{\sun}$) systems, have been compared to the local
SDSS relation measured by \cite{Shen2003} for both red and blue,
concentrated and diffuse systems. 
The results to-date provide an intriguing yet consistent picture of significant size growth
from z $>$ 1.5 to z=0.0 with minimal mass
increase (e.g. \citealt{Ferguson2004, Daddi2005, Longhetti2007,
  VanDokkum2008, Trujillo2006}). 
 These initial results have been corroborated by extensive studies which
continue to identify a clear disconnect between the \msr relation
of nearby galaxies and those at intermediate- to high-redshift \citep{Trujillo2007, Buitrago2008, Wel2008, Damjanov2009, McIntosh2005, Williams2010, Bruce2012}. 
The current data, mostly confined to massive
early-type systems, seem to suggest that galaxies have grown by a
factor of five in size since z$\sim$2 with minimal change in mass. 
By contrast, the \msr evolution of disk systems is traced at lower redshift (z$\lesssim$1) and
appears less dramatic, evolving by roughly a factor of 2 (see for example
\citealt{Barden2005, Sargent2007, Buitrago2008, VanDokkum2013}). 
A number of physical and non-physical explanations have been put forward to explain the observed \msr evolution of the early-types.
These include, for example, major and minor mergers or gas accretion and disc growth as physical effects, see e.g.~\cite{Driver2013} and also \cite{Graham2013} who suggest that the compact galaxies at high-z are the naked bulges of lower-z systems. 
Some evidence for this scenario is suggested by the compact massive bulges evident in nearby early-type galaxies seen by Dullo \& Graham (2013). Non-physical and systematic effects may include various selection biases, as well as erroneous estimations of mass and size (see e.g.~\citealt{Hopkins2009} and references therein). \\

 Finally it is important to note that the often used redshift zero \msr relation of \citet{Shen2003} measures sizes in the \textit{z}-band and uses a S\'ersic index cut to divide the galaxy sample into early- and late-types. 
 We have already discussed the wavelength dependent size of galaxies, but another caveat is the definition of early-\slash{}late-type. Commonly colour, concentration (i.e.~S\'ersic index) or morphology are used interchangeably, but even though there is a correlation these definitions are not synonymous \citep{Robotham2013}, for example low-luminosity elliptical galaxies can have S\'ersic indices of n$<$2.5 (e.g. \citealt{Graham2003} and references therein).
Hence, when comparing the local \msr relation to other data sets due consideration should be given to the necessary correction of the wavelength dependent sizes of galaxies and the method used to separate the sample into early- and late-type. \\

In this paper we provide a comprehensive recalibration of the local \msr relation, divided into early- and late-type galaxies, according to various criteria which include: S\'ersic index, colour, a joint S\'ersic index-colour cut and galaxy visual morphology. Due to the similarity between the morphology-dependent mass- size relation and the fundamental mass-spin-morphology relation \citep{Cappellari2011,Romanowsky2012,Obreschkow2014}, this work lays the foundations for approximate studies of angular momentum scalings in a large local sample with well-characterized completeness. We will expand this idea in sequel work. 
In addition, for comparison between different redshifts, we derive the \msr relation in a consistent manner for 10 imaging bands (\textit{ugrizZYJHK$_s$}). \\

 Throughout this paper we use data derived from the Galaxy And Mass Assembly (GAMA) survey \citep{Driver2011,Liske2014} with stellar mass estimates as described in \cite{Taylor2011}, half-light radii derived from 2D S\'ersic light profile fitting as described in \cite{Kelvin2012} and for a cosmology given by a $\Lambda{}$CDM universe with:\\ 
$\Omega_{\mathrm{m}}=0.3,\; \Omega_{\Lambda}=0.7$ and $H_0=70\mathrm{kms}^{-1}\;\mathrm{Mpc}^{-1}$.

\section{Data}
\label{sec:data}

In this section we briefly describe the GAMA data (section
\ref{subsec:survey}), the derived stellar masses
(\ref{subsec:mass}), galaxy sizes (\ref{subsec:size}) and 
the sample selection (\ref{sec:selection}) used in this
paper.

\subsection{The GAMA survey}
\label{subsec:survey}
The GAMA survey is an optical spectroscopic and multi-wavelength
imaging survey combining the data of several ground and space based
telescopes \citep{Driver2011}. It is an
intermediate survey in respect to depth and survey area (see
\citealt{Baldry2010}; figure 1) and thus fits in between low
redshift, wide-field surveys such as SDSS \citep{York2000} or 2dFGRS
\citep{Colless2003} and narrow deep field surveys like zCOSMOS
(\citealt{Lilly2007} and see \citealt{Davies2014}) or DEEP-2 \citep{Davis2003}. \\

In this paper we are selecting data from the \mbox{GAMA II} \citep[see the second data release paper,][]{Liske2014} equatorial regions, which are centered on 9h (G09), 12h (G12) and
14.5h (G15).
The three regions are 12$\times$5 $\mathrm{deg}^{2}$ and have a \textit{r}-band Petrosian magnitude limit of $r<$19.8 mag.
The spectroscopic target selection is derived from 
a SDSS DR 7 \citep{Abazajian2009} input catalogue and we reach a spectroscopic completeness of
$\ge$ 98\% for the main survey targets.
The available survey bands include SDSS DR7 (\textit{ugriz} bands), UKIDSS LAS DR6 and DR8 (\textit{YJHK} bands, \citealt{Lawrence2007}  ) and VISTA (Visible and Infrared Telescope for Astronomy) Kilo-degree INfrared Galaxy survey (VIKING) data (\textit{ZYJHK$_s$} bands, \citealt{Edge2013} and also see Driver et al.~in prep.~for more details on the GAMA processing of the VIKING data).
All imaging data has matched aperture photometry \citep{Hill2011, Liske2014} 
and the spectroscopic redshifts \citep{Baldry2014, Liske2014} are based on spectra 
taken with AAOmega (resolution of R$\sim$1300) 
at the 3.9m Anglo-Australian-Telescope \citep{Hopkins2013} located at Siding Spring Observatory (NSW, Australia).

\subsection{Stellar Masses}
\label{subsec:mass}

The stellar mass estimates for GAMA are described in \citet{Taylor2011}
and are based on synthetic stellar population models from the
BC03 library \citep{BC03} with a \citet{Chabrier2003} initial mass function and 
the \citet{Calzetti2000} dust obscuration law. 

The stellar masses are estimated from the best fitting broadband spectral energy distributions (SEDs), which are generated using stellar population synthesis modelling and compared to observed GAMA SEDs  in a fixed restframe wavelength range from 3000 to 11000 \AA\  (roughly \textit{u} to\textit{ J}-band, depending on the redshift of the source).

It is important to note that no further near-infrared (NIR) photometry is used for the stellar mass estimates.
The colour-colour space in the NIR can not be adequately sampled with little present metallicity which makes the modelling of the NIR SEDs difficult. In addition, there may be a problem with the NIR data which in the original analysis \citet{Taylor2011} led to the exclusion of the entire available NIR data. 

Here we are using the stellar masses v16 catalogue and the stellar masses, 
based on aperture matched photometry, are believed
to be accurate to within a factor of 2.

Additionally we apply the  `fluxscale' parameter, given in the catalogue, to our masses to correct for aperture sizes. Since the GAMA SEDs are derived from matched aperture photometry, which is based on the \emph{SExtractor} AUTO magnitudes, integrated quantities such as the stellar mass need an aperture correction to account for the mass that lies outside the fixed AUTO aperture. 
The fluxscale parameter is the ratio between the \textit{r}-band AUTO flux and the total flux of a source derived from its 10R$_e$ truncated S\'ersic profile.

\begin{figure}
\includegraphics[width=0.45\textwidth]{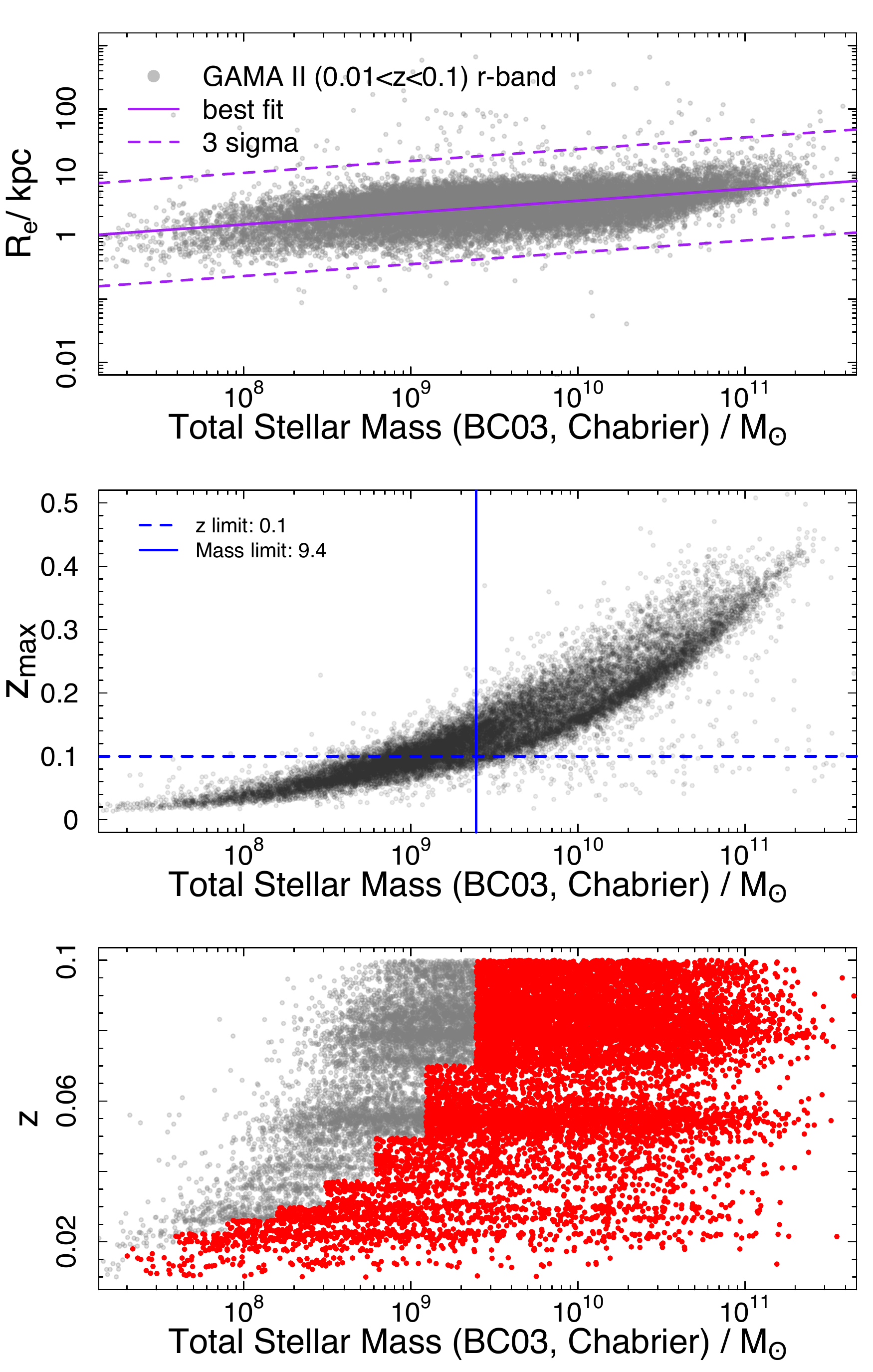}
\caption{(Top) The stellar mass -- (major axis) half-light radius 
distribution of the sample extracted from the
  GAMA catalogue with all galaxies in the redshift range between
  0.01$<$z$<$0.1 as grey dots, the solid purple line shows an initial least square fit to the
entire data set and the dashed lines indicate the 3$\sigma$
  scatter. Galaxies more than 3$\sigma$ away from the best fit are excluded from
  the final data set. \newline (Middle) The stellar mass
  distribution vs maximum redshift ($z_{\rm max}$) at which the galaxy can be seen for the limiting petrosian magnitude of $r=19.8$ mag. 
  The black points show the GAMAmid sample, the dashed line indicates the adopted upper redshift limit of the sample (z=0.1) and the solid line shows the calculated mass limit for which 97.7\% of the galaxies
  have a $z_{max} $ above the indicated redshift limit. \newline (Bottom) the stellar mass distribution vs redshift for the GAMAmid sample shown in grey and the staggered volume limited sample highlighted in red. Each mass bin has an associated weight that is used to weight each galaxy within the respective bin.}
\label{fig:diagnostics}
\end{figure}

\subsection{Galaxy Sizes and S\'ersic Index}
\label{subsec:size}

The galaxy sizes (i.e. the effective major axis half-light radius) are based on
single S\'ersic 2D model fits to the data in 10 bands (\textit{ugrizZYJHK$_s$}, see \citealt{Kelvin2012} for details on the fitting pipeline). 
The original S\'ersic profile fitting used imaging data obtained from SDSS DR7 and UKIDSS LAS, 
which were reprocessed and scaled to a single zero point and then mosaiced with SWARP \citep{Bertin2002} at a resolution of 0.339$''$ (see \citealt{Hill2011} and Driver et al.~in prep.). 
The VIKING data is handled in a similar way to the UKIDSS data, i.e. scaled to the same zero point and `swarped' with a pixel resolution of 0.339$''$ (Driver et al.~in prep.). 
The mosaics along with the GAMA input catalogue are fed into SIGMA (Structural
Investigation of Galaxies via Model Analysis, \citealt{Kelvin2012}) an
automated front-end wrapper which uses a range of image analysis
software (such as Source Extractor, \citealt{Bertin1996}; PSF
Extractor, \citealt{BertinPSF} and GALFIT3, \citealt{Peng2010}), as
well as logical filters and other handlers to carry out bulk analysis
on the input catalogue.

The final output of SIGMA provides values for S\'ersic index,
effective half-light radius, position angle, ellipticity and magnitude (defined
according to the AB magnitude system).  
Here we are using the pre-release of version 9 of the S\'ersic fits catalogue and we have opted to use the VIKING \textit{ZYJHK$_s$} fitting results instead of the UKIDSS \textit{YJHK} results.
 The improved imaging quality of the VIKING data allows for more robust S\'ersic light profile fitting \citep{Andrews2014}, which in turn means that our \msr relation fits in the \textit{ZYJHK$_s$} bands are also more robust.

\subsection{Sample Selection}
\label{sec:selection}

 In this work we selected galaxies from the GAMA equatorial regions in the redshift range of
0.01 $\leq$ z $\leq$ 0.1 with redshift qualities nQ$\ge$ 3\footnote{Spectra with a nQ flag of 3 and higher have good quality redshifts with probabilities $p(z)>0.9$ and can be used for scientific analysis \citep{Liske2014}.}, vis$_-$class$!=$3\footnote{Sources with vis$_-$class=3 are classed as `not a target' since they are not the main part of a galaxy.} and magnitudes r$<$ 19.8 mag for G09, G12 and G15.\

The top panel in Fig.~\ref{fig:diagnostics} shows the distribution of the half-light
radius vs the galaxy stellar masses for the entire sample of
20287 objects in the \textit{r}-band, the solid line is a least square fit to the
data and the dashed lines indicates the 3$\sigma$ spread; 
outliers are defined as being more than 3$\sigma$ from the
best fit.  A visual inspection of all 241 outliers showed
that most of these galaxies were close to bright stars which contaminated the flux measurements, consequently these galaxies were removed from the sample.
In addition galaxies with unrealistic fitting parameters such as S\'ersic indices (n$\leq$0.3 or n$\geq$10) and sizes (R$_{e}<$0.5$\times$FWHM) were also excluded from the sample. After the exclusion of the outliers, unrealistic and failed fits the \textit{r}-band `good fit' sample consists of 18795 galaxies and is referred to as GAMAmid hereafter. All other bands are treated the same way to establish the `good fit' sample which is shown in Table \ref{table:samplesize}. \\

For each galaxy in our sample \cite{Taylor2011} calculated the maximum
redshift (z$_{max}$) to which this object could be detected given its
best fit spectral template and an apparent \textit{r}-band Petrosian magnitude
of 19.8 mag, the limiting magnitude of the GAMA-II data release.  
To establish the lower mass limit for a volume limited sample we check which galaxies are visible at or beyond the adopted upper redshift limit (i.e.~z$_{max}>$0.1).

The middle panel of Fig.~\ref{fig:diagnostics} shows the stellar mass
distribution versus $z_{max}$ 
based on each galaxy's spectral
shape and our \textit{r}-band magnitude limit.  The blue dashed horizontal line shows the
redshift limit of z=0.1 and the solid vertical line indicates the
lower mass limit for the sample set at the 97.7\% level, i.e. of all
the galaxies above the mass limit 97.7\% can be seen at or beyond the chosen
redshift limit. 
This results in a lower mass limit of {$\mathcal{M}_{lim}$=2.5$\times\mathrm{10}^{9}\mathcal{M}_{\sun}$} to ensure a colour-unbiased sample of 9751 galaxies.

However, using this mass limit means we would discard $\sim$50\% of our data, so in order to include lower mass galaxies we use a staggered volume \& mass limited selection.
 To implement this staggered limit we divide the galaxies below our mass limit into bins with size $\Delta$log$_{10}$(M$_*$)=0.3. For each bin we establish the expected maximum redshift at the lower mass end (z$_{bin}$) which satisfies the completeness criterion. We then discard all galaxies with redshifts z$>$z$_{bin}$, the results can be seen in the bottom panel of  Fig.~\ref{fig:diagnostics}.\\
 
The  galaxies remaining within the bin are equally weighted by a common weight W$_{bin}$ which is based on a V/V$_{max}$ of z$_{bin}$.\\
V/V$_{max}$ is calculated by computing the ratio of the volume in which a galaxy is seen over the maximum volume in which the galaxy can be seen:
\begin{equation}
\label{equ:v-vmax}
 \mathrm{W_{bin}=\frac{V(z)} {V(z_{max})}}.
\end{equation}

Here we calculate V(z) using the redshift assigned to each bin and V$_{max}$ by setting the maximum redshift to be z$_{max}$=z$_{lim}$=0.1. For the low mass galaxies in our sample with  M$< M_{lim}$ we weight each galaxy according to the corresponding weight of the bin W$_{bin}$ and for galaxies with M$\geq M_{lim}$ the weight is set to 1. This ensures that all galaxies within the staggered volume limited sample get up-weighted and galaxies within the unbiased volume limited sample are not penalized. Furthermore using the staggered volume limited sample ensures that no single galaxy will overly influence the fitting routine because of a very large individual weight.\footnote{An individual V/V$_{max}$ based on each galaxy's z$_{max}$ can cause a few data points to skew the \msr relation. We found this to be especially problematic in the case of the S\'ersic cut early-type galaxies.}\\
 
Treating each band in this way leads to similar volume limited sample sizes ($<$8\% difference in samples), which confirms we observe essentially the same galaxy populations in each waveband. However, to ensure that we do not introduce any biases even within these small fluctuations we have decided to establish a common set. This sample includes only those galaxies from our volume limited sample that have good S\'ersic profile fitting parameters in all bands except \textit{u} (which is not considered here due to its poor imaging quality). This reduces our final sample to 8399 galaxies, which is used to fit the \msr relation from \textit{g}-band to K$_s$-band. 

We additionally set up a second common sample which includes the \textit{u}-band data, which reduces the final sample size to 6154 galaxies. This sample is only used to fit the \msr relation in the \textit{u}-band.  We do this to ensure all the other bands are not penalized for the bad image quality in the \textit{u}-band. Hence, we do not include the \textit{u}-band \msr relation fits in subsequent comparisons but present the results in Tables \ref{table:rmfitsL} and \ref{table:rmfitsE} for completeness. 

\begin{table*}
\centering
\begin{tabular}{|l|l|l|l}\\
\\
	\hline
		Band & \multicolumn{3}{c}{0.01$\leq$z$\leq$0.1 sample size} \\
	&  volume limited &		colour unbiased &	 staggered volume limited \\ 
	\hline \hline
	\\
\textit{u}     		& 		10830 		& 		6904		&		8343 \\
\textit{g}     		& 		18321 		& 		9555		&		11813 \\
\textit{r}			& 		18795		&		9751		&		12037\\
\textit{i}      		&	 	18445 		& 		9619		&		11887 \\
\textit{z}     		&		15558 		& 		9227		&		11193 \\
Z   					&		18214 		& 		9373		&		11602 \\
Y     					& 		18140 		&		9411		&		11621 \\
J     					& 		18764 		& 		9730		&		11993 \\
H     					& 		17626 		& 		9296		&		11449 \\
K$_s$     			& 		17790 		& 		9434		&		11581 \\
common excl.~u & - 	& 		-				&		8399 \\
common incl.~u & - 	& 		-				&		6154 \\
\hline
\end{tabular}
\caption{From left to right we show the volume limited sample size in each band after outliers, bad and failed fits have been removed; the colour unbiased sample size (i.e.~number of galaxies above the mass limit) and the final sample size after the staggered volume limited selection is implemented. The last two rows show the final common sample, excluding and including the \textit{u}-band respectively. These are based on all galaxies common in all bands within the staggered volume limited samples and are used for the \msr relation analysis.}
\label{table:samplesize}
\end{table*}

\section{\msr Relations by Early- and Late-Type}
\label{sec:MSR}

\begin{figure}
\centering
\includegraphics[width=0.49\textwidth]{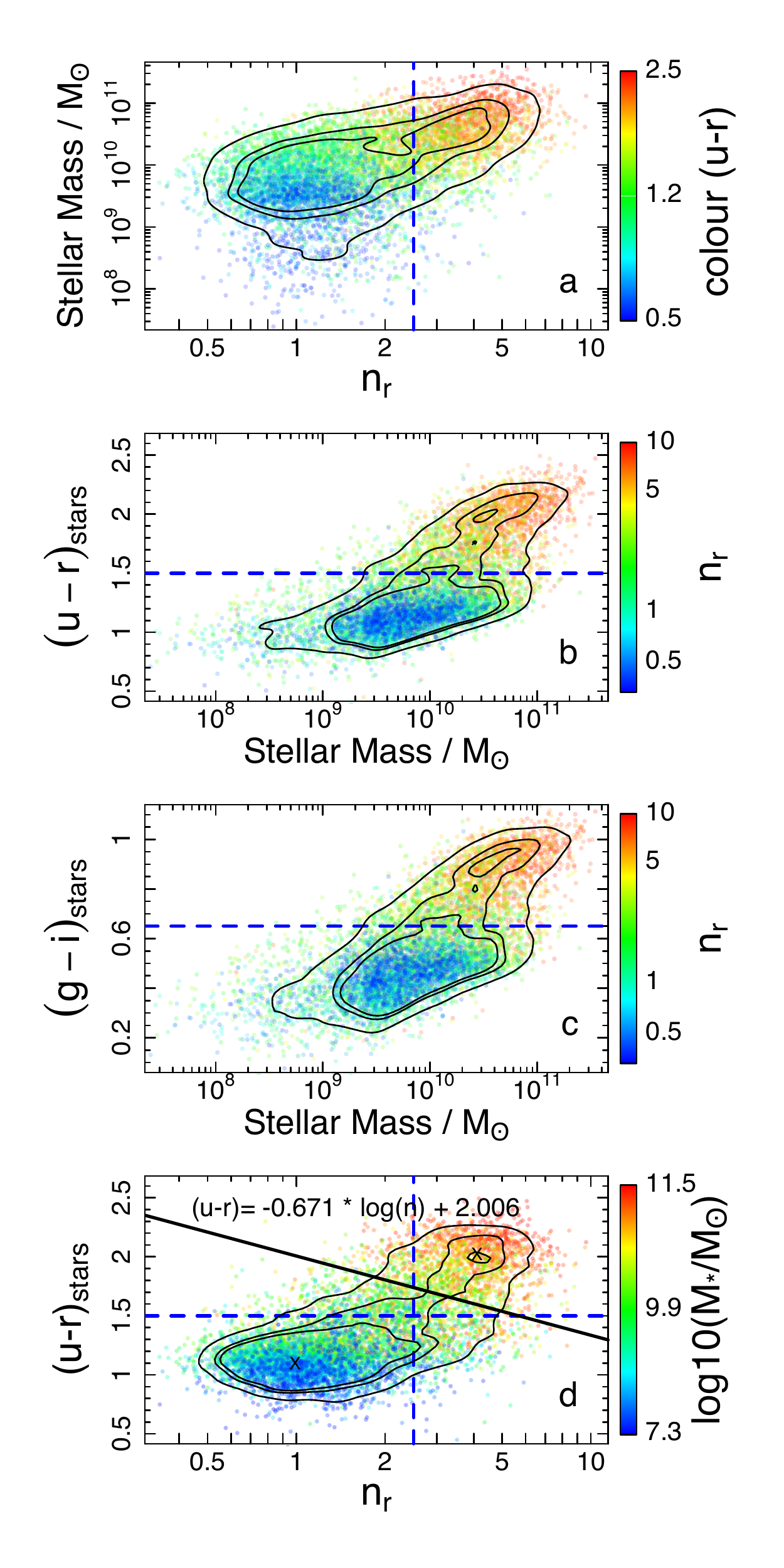}
\caption{The plot shows the sample distribution in a 3D parameter space
illustrating the population selection criteria adopted for the
\msr relation.\newline (a) The S\'ersic index v total stellar
mass, colour coded by (\textit{u-r})$_{stars}$ colour.
\newline (b) The (\textit{u-r})$_{stars}$ colour v
total stellar mass, colour coded by S\'ersic index. 
\newline (c) The (\textit{g-i})$_{stars}$ colour v
total stellar mass, colour coded by S\'ersic index.
\newline (d) The S\'ersic index v (\textit{u-r})$_{stars}$ colour, colour coded 
by total stellar mass.
\newline The blue dashed lines show the hard cuts adopted for
S\'ersic index (n=2.5) and colour (u-r=1.5) and the solid black line
in the bottom panel is a combined S\'ersic index and colour cut which 
gives the best population division (in respect to the visual classifications) 
with (\textit{u-r})$_{stars}=-0.671\times\log_{10}(\rm n_r)+2.006$.}
\label{fig:select}
\end{figure}

In this section we derive \msr relations as a function of galaxy type. 
For this we divide the GAMAmid common sample into early- and
late-types (see Fig.~\ref{fig:select}) according to the S\'ersic index n (see
Section \ref{sec:sersic}), the dust corrected restframe (\textit{u-r})$_{stars}$ and (\textit{g-i})$_{stars}$ colours (Section \ref{sec:colour}), a combined S\'ersic index and (\textit{u-r})$_{stars}$ colour division (Section \ref{sec:Roll}) and galaxy visual morphology (Section \ref{sec:morph}).
In each section early- and late-type is defined by the chosen separator and we strongly caution that this is not to be confused with actual elliptical or disc galaxy populations, except for Section \ref{sec:morph} in which we split the population by visual morphology.\\

We fit all early- and late-type \msr relations using two functions motivated by \cite{Shen2003} (S03 hereafter) in order to directly compare with their work.

 Firstly a single power-law function:

\begin{equation}
\label{equ:rm1}
R_e=a \mathrm{\left(\frac{\mathcal{M}_*} {\mathcal{M}_{\sun}}\right)}^{b},
\end{equation}

 and secondly a combination of two power-law functions:

\begin{equation}
\label{equ:rm2}
R_e={\gamma} \mathrm{\left(\frac{\mathcal{M}_*} {\mathcal{M}_{\sun}}\right)}^{\alpha} {\left(1+\frac{\mathcal{M}_*} {\mathcal{M}_0}\right)}^{\beta - \alpha} ,
\end{equation}

where R$_e$ is the effective half-light radius in kpc, $\mathcal{M}_*$ is the mass of the
galaxy and M$_0$ (the breakpoint between the two power-law functions) can be 
considered an artificial transition mass between low-
and high-mass galaxies (in units of $\mathcal{M}_{\sun}$) in any given sample. 

We use Bayesian inference with an MCMC approach to find the expectation of parameters describing the data.
For this we weight each data point by the V/V$_{max}$ which is associated with the mass bin in which the data point lies (see previous section explaining the staggered volume limited sample) and use uniform priors to perform our fitting. 

Except for an upper limit on M$_0<10^{13}\mathcal{M}_{\sun}$, we do not restrict any parameters in Eqs.~\ref{equ:rm1} and \ref{equ:rm2} during the fitting process and caution that the resulting regression lines should be considered (if possible) only within the mass range for which they were fit. The fitting is performed on the entire sample and median data points shown in our figures are for visualisation only (Figure \ref{fig:rmall}, as well as the figures in the appendix).\\

We have also calculated the regions in which our data becomes less reliable and show these as shaded areas in the \msr relation plots (Figs.~\ref{fig:rmall} and \ref{fig:msru}-\ref{fig:msrk}). In total we define the three boundaries (following \citealt{Driver1999}):
\begin{enumerate}
\item The minimum size boundary\\
This area indicates where the star-galaxy separation becomes difficult since the galaxies are only marginally resolved, i.e. they have R$_{e}<$0.5$\times$FWHM. Note that the lower boundary we plot shows the typical \textit{r}-band size limit expected for the redshift (z$_{max}$) in each mass bin using the average SDSS FWHM of 1.5$''$ to calculate the equivalent radius in kpc. Please also note that this is not a hard lower limit and we check for each galaxy if its R$_e$ is smaller than the FWHM of its image frame, this leads to galaxies being found within the (average) minimum size boundary.\\

\item The maximum size boundary\\
Due to the way sky subtraction and background noise is handled, galaxies that are very large run the risk of contributing to the sky background estimation and hence their sizes become questionable.
This becomes a problem when a galaxy occupies 20\% of the pixels within the background sampling box\footnote{Initial background subtraction is performed during{}\begin{tiny} SWARP\end{tiny} using a 256 x 256  pixel mesh \citep{Driver2011}.} and in our case equates to a FWHM=20$''$.
The corresponding size is calculated in kpc for all redshift bins. However, due to surface brightness considerations the maximum size boundary only comes into effect for very high mass galaxies.\\

\item Surface brightness boundary\\
Considering the \textit{r}-band surface brightness (24.5mag\slash{}arcsec$^2$) and magnitude limit (19.8 mag) of the survey,  we can derive an upper boundary at which galaxies become too large to be easily detected (i.e $\mu_{\rm eff}\sim\mu_{\rm lim}$).\\

First we consider the surface brightness:
\begin{equation}
\label{equ:sb}
\mu_{\rm eff}=m + 2.5 \log_{10}(2\pi\theta^2),
\end{equation}
where $\mu_{\rm eff}$ is the effective surface brightness, $m$ is the apparent magnitude and $\theta$ is the angular size.\\

\noindent Then we need to consider the apparent magnitude:
\begin{equation}
\label{equ:mag}
m={M_*} + 5 \log_{10}(d_l) + 25 + k(z),
\end{equation}
where $M_*$ is the absolute magnitude, $d_l$ is the luminosity distance in Mpc and $k(z)$ is the K-correction.\\

Relating the absolute magnitude to solar units we find:
\begin{equation}
\label{equ:magsol}
{M_*-M_{\sun}}= - 2.5 \log_{10}\left(\frac{L_*} {L_{\sun}}\right)= - 2.5 \log_{10}\left(\frac{\frac{L_*} {\mathcal{M_*}} \mathcal{M_*}} {\frac{L_{\sun}}{\mathcal{M}_{\sun}}\mathcal{M}_{\sun}}\right),
\end{equation}
where $M_{\sun}$ is the absolute magnitude of the sun, $L_*$ and $L_{\sun}$ are the luminosity of the galaxy and the sun respectively and $\mathcal{M}_{*}$ and $\mathcal{M}_{\sun}$ are the corresponding masses.

Re-arranging Eq.~\ref{equ:magsol} and substituting it into Eqs.~\ref{equ:mag} and \ref{equ:sb} we can derive an upper size limit for our redshift bins using the surface brightness and magnitude limits of the GAMA survey: 
\begin{equation}
\label{equ:rlim}
\theta = \sqrt{\frac{\frac{L_*} {\mathcal{M_*}} \mathcal{M}} {2\pi}} \frac{1}{d_l} 10^{0.2(\mu_{\rm lim}-M_{\sun}-k(z)-25)}
\end{equation}
where $\mathcal{M}$ is the galaxy mass in units of $\mathcal{M}_{\sun}$ and we assume an \textit{i}-band $\frac{\mathcal{M_*}} {L_*}$=2 \citep{Baldry2010}, $M_{\sun}=4.6$ \citep{Hill2010} and $k(z)=1.5z$ \citep{Driver1994}.

 The angle is converted to a physical size for each mass bin by considering the lower and upper mass boundaries of the bin and its corresponding redshift limit.
\end{enumerate}

Note that the boundaries are not strict limits but represent the regions where measurements become less robust. We find that, while these boundaries enclose our data, they do not shepherd it (see Fig.~\ref{fig:rmall}) as a fall off in the density of data points is seen before the boundaries are encountered. We therefore conclude that the \msr relations are not being led by the selection boundaries. \\

We have chosen the \textit{r}-band to present our method  since it is the spectroscopic selection band for the GAMA survey and is also a commonly used band in other studies. However, we have fit all bands (\textit{ugrizZYJHK$_s$}) and the results are presented alongside the \textit{r}-band parameters in Tables \ref{table:rmfitsL} and \ref{table:rmfitsE} and are plotted in the appendix.

\subsection{\msr Relation: division by S\'ersic index}
\label{sec:sersic}

We first compare the \msr relation of our sample with the
relation found by \citet{Shen2003} (see their Fig.~11)
for high and low S\'ersic index selected samples. We then go on to
discuss other possible S\'ersic population separators currently in
use.\\

The S\'ersic profile \citep{Sersic1963,Sersic1968,Graham2005} describes a galaxy's intensity, $I(r)$, as a function of radius, $r$:
\begin{equation}
I(r)=I_e \exp \left[-b_n \left( \left(\frac{r}{r_e} \right) ^{1/n} -1 \right) \right],
\end{equation} 
where $I_e$ is the intensity at the effective radius $r_e$, i.e. the half-light radius. The parameter $b_n$ is a function of the S\'ersic index n, such that $\Gamma(2n) = 2 \gamma(2n,b_n)$, where $\Gamma$ and $\gamma$ are the complete and incomplete gamma functions respectively \citep{Ciotti1991}. The S\'ersic index, n, describes the shape of the
light profile, such that n=0.5 gives a Gaussian profile, n=1 describes an exponential profile and n=4 recovers the de Vaucouleurs $r^{1/4}$ light profile. The S\'ersic index can also be thought of as a concentration index of the galaxy \citep{Trujillo2001} where high S\'ersic index galaxies are more centrally concentrated than low S\'ersic index galaxies.

\subsubsection{Comparison with S03 - S\'ersic index n=2.5}

\begin{figure*}
\begin{minipage}[c][\textheight]{\textwidth}
\centering
\includegraphics[width=0.87\textwidth] {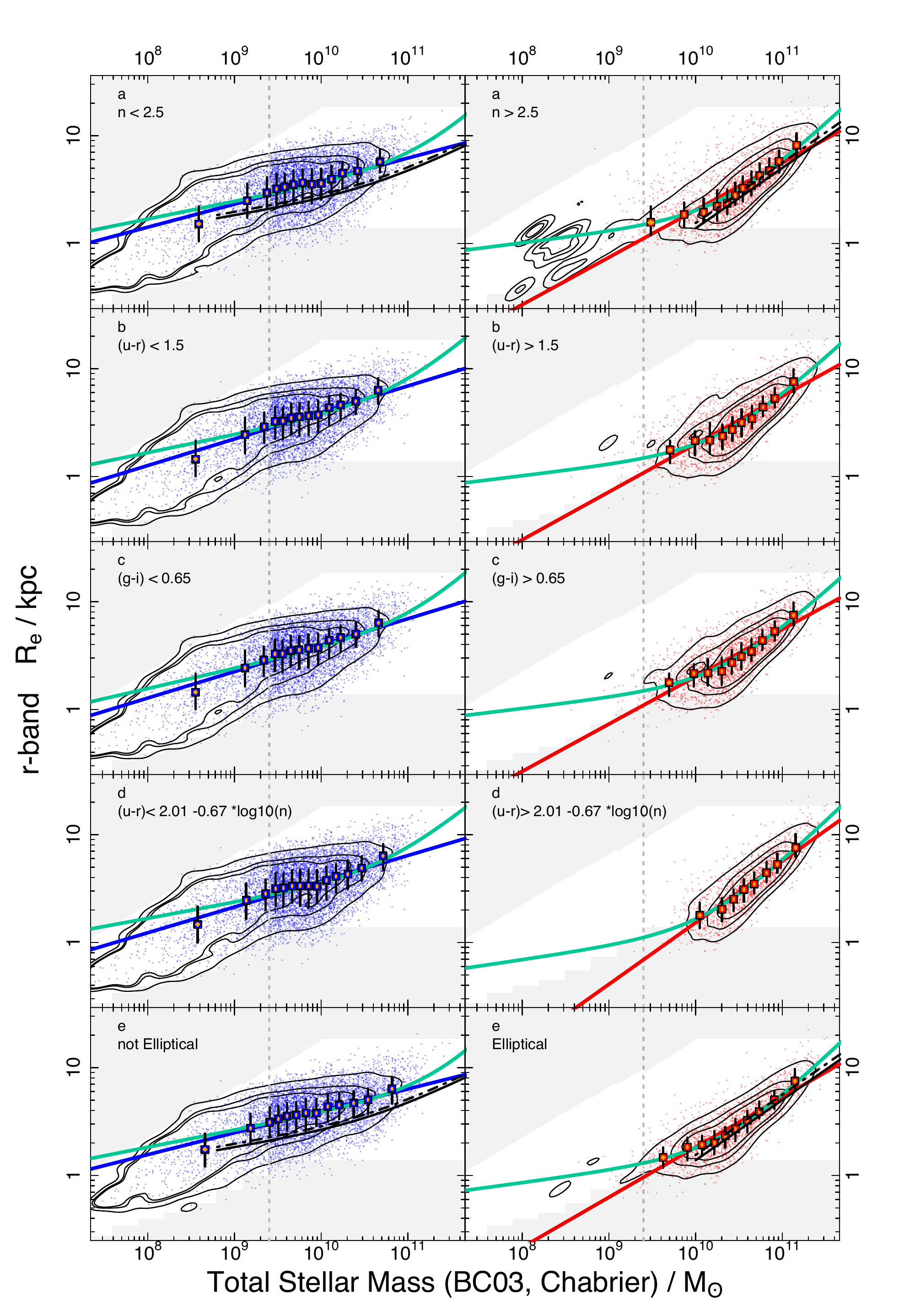}
\caption{The \msr relation for early- (red, right hand side) and late- (blue, left hand side) type galaxies
divided by:\newline 
a) S\'ersic index n=2.5; 
b) Dust corrected colour (\textit{u-r})$_{stars}$=1.5;
c) Dust corrected colour (\textit{g-i})$_{stars}$=0.65;
d) Rolling S\'ersic index division and
e) Visual Elliptical/Not-Elliptical classification. 
The red and blue lines are single power-law fits to the binned data
(Eq.~\ref{equ:rm1}), the green lines are two component power-law fits
(Eq.~\ref{equ:rm2}) and the grey dotted line indicates the lower mass
limit highlighting the wealth of data that would have been ignored. The grey shaded areas indicate where measurements become less reliable due to our detection limitations.
The black solid and dot-dashed lines in panels a) and e) show the S03 \msr relation where the dot-dashed line shows the sizes corrected from z- to \textit{r}-band values and the solid line shows the relation as is.
For the fitting parameters see Tables \ref{table:rmfitsL} and \ref{table:rmfitsE}.}
\label{fig:rmall}
\end{minipage}
\end{figure*}

\begin{table*}
\centering
\begin{tabular}{|l|l|l|llll|ll|l|}\\
\\
	\hline
	\multicolumn{3}{l}{\textbf{Late-type galaxies}}&&\\
	\\
Case	&& a{ } ($\mathrm {10}^{-3}$) & b && ${\alpha}$ & ${\beta}$ &$ {\gamma} $& M$_0${ } ($\mathrm {10}^{10}\mathcal{M_{\sun}}$) \\ 
	\hline \hline
	\multicolumn{3}{l}{\textbf{S\'ersic cut}} \\
u (Fig.~\ref{fig:msru})  && 74.46 $\pm$ 15.16  & 0.17 $\pm$ 0.02  && 0.12 $\pm$ 0.04 & 0.88 $\pm$ 0.39 & 0.21 $\pm$ 0.12 & 27.18 $\pm$ 1.56\\
g (Fig.~\ref{fig:msrg})  && 24.48 $\pm$ 3.07  & 0.22 $\pm$ 0.02  && 0.16 $\pm$ 0.04 & 0.78 $\pm$ 0.30 & 0.08 $\pm$ 0.03 & 15.77 $\pm$ 0.82\\
r (Fig.~\ref{fig:rmall})  && 27.72 $\pm$ 3.93  & 0.21 $\pm$ 0.02  && 0.16 $\pm$ 0.04 & 0.81 $\pm$ 0.32 & 0.08 $\pm$ 0.03 & 17.10 $\pm$ 0.91\\
i (Fig.~\ref{fig:msri})  && 23.36 $\pm$ 3.18  & 0.22 $\pm$ 0.02  && 0.16 $\pm$ 0.04 & 0.76 $\pm$ 0.29 & 0.09 $\pm$ 0.03 & 11.23 $\pm$ 0.56\\
z (Fig.~\ref{fig:msrz})  && 35.37 $\pm$ 5.69  & 0.20 $\pm$ 0.02  && 0.15 $\pm$ 0.04 & 0.87 $\pm$ 0.36 & 0.11 $\pm$ 0.04 & 17.71 $\pm$ 1.00\\
Z (Fig.~\ref{fig:msrzv})  && 34.29 $\pm$ 4.88  & 0.20 $\pm$ 0.02  && 0.15 $\pm$ 0.04 & 0.84 $\pm$ 0.34 & 0.10 $\pm$ 0.03 & 19.23 $\pm$ 1.01\\
Y (Fig.~\ref{fig:msry})  && 28.52 $\pm$ 4.19  & 0.21 $\pm$ 0.02  && 0.16 $\pm$ 0.04 & 0.81 $\pm$ 0.33 & 0.09 $\pm$ 0.03 & 15.60 $\pm$ 0.86\\
J (Fig.~\ref{fig:msrj})  && 28.69 $\pm$ 4.46  & 0.21 $\pm$ 0.02  && 0.16 $\pm$ 0.04 & 0.86 $\pm$ 0.36 & 0.08 $\pm$ 0.03 & 17.97 $\pm$ 1.01\\
H (Fig.~\ref{fig:msrh})  && 25.26 $\pm$ 3.91  & 0.21 $\pm$ 0.02  && 0.17 $\pm$ 0.04 & 0.88 $\pm$ 0.38 & 0.07 $\pm$ 0.02 & 20.14 $\pm$ 1.27\\
K (Fig.~\ref{fig:msrk})  && 27.19 $\pm$ 4.50  & 0.21 $\pm$ 0.03  && 0.17 $\pm$ 0.04 & 0.94 $\pm$ 0.40 & 0.06 $\pm$ 0.02 & 26.37 $\pm$ 1.74\\
				S03 									&& -	& - 						&& 0.14 & 0.39 & 0.1 	& 3.98 \\
	\hline	
\multicolumn{3}{l}{	\textbf{(\textit{u-r}) colour cut } }\\
u (Fig.~\ref{fig:msru})  && 16.67 $\pm$ 2.40  & 0.24 $\pm$ 0.02  && 0.16 $\pm$ 0.05 & 0.77 $\pm$ 0.29 & 0.09 $\pm$ 0.04 & 9.82 $\pm$ 0.52\\
g (Fig.~\ref{fig:msrg})  && 11.79 $\pm$ 1.24  & 0.25 $\pm$ 0.02  && 0.17 $\pm$ 0.05 & 0.72 $\pm$ 0.26 & 0.07 $\pm$ 0.03 & 7.66 $\pm$ 0.43\\
r (Fig.~\ref{fig:rmall})  && 13.63 $\pm$ 1.65  & 0.25 $\pm$ 0.02  && 0.17 $\pm$ 0.04 & 0.73 $\pm$ 0.26 & 0.08 $\pm$ 0.03 & 8.56 $\pm$ 0.45\\
i (Fig.~\ref{fig:msri})  && 11.79 $\pm$ 1.34  & 0.25 $\pm$ 0.02  && 0.16 $\pm$ 0.05 & 0.67 $\pm$ 0.20 & 0.09 $\pm$ 0.04 & 5.76 $\pm$ 0.23\\
z (Fig.~\ref{fig:msrz})  && 15.86 $\pm$ 1.95  & 0.24 $\pm$ 0.02  && 0.15 $\pm$ 0.04 & 0.71 $\pm$ 0.22 & 0.10 $\pm$ 0.04 & 7.02 $\pm$ 0.29\\
Z (Fig.~\ref{fig:msrzv})  && 24.77 $\pm$ 3.32  & 0.22 $\pm$ 0.02  && 0.16 $\pm$ 0.04 & 0.83 $\pm$ 0.34 & 0.08 $\pm$ 0.03 & 16.02 $\pm$ 0.90\\
Y (Fig.~\ref{fig:msry})  && 19.59 $\pm$ 2.41  & 0.23 $\pm$ 0.02  && 0.15 $\pm$ 0.04 & 0.75 $\pm$ 0.28 & 0.10 $\pm$ 0.04 & 9.98 $\pm$ 0.49\\
J (Fig.~\ref{fig:msrj})  && 19.44 $\pm$ 2.51  & 0.23 $\pm$ 0.02  && 0.15 $\pm$ 0.04 & 0.74 $\pm$ 0.26 & 0.10 $\pm$ 0.05 & 8.80 $\pm$ 0.39\\
H (Fig.~\ref{fig:msrh})  && 15.50 $\pm$ 1.80  & 0.23 $\pm$ 0.02  && 0.15 $\pm$ 0.04 & 0.71 $\pm$ 0.22 & 0.10 $\pm$ 0.04 & 7.18 $\pm$ 0.29\\
K (Fig.~\ref{fig:msrk})  && 11.12 $\pm$ 1.26  & 0.25 $\pm$ 0.02  && 0.15 $\pm$ 0.05 & 0.68 $\pm$ 0.18 & 0.10 $\pm$ 0.05 & 5.09 $\pm$ 0.18\\
	
	\hline	
\multicolumn{3}{l}{	\textbf{(\textit{g-i}) colour cut } }\\
u (Fig.~\ref{fig:msru}) && 16.89 $\pm$ 2.42  & 0.24 $\pm$ 0.02  && 0.16 $\pm$ 0.05 & 0.79 $\pm$ 0.29 & 0.10 $\pm$ 0.05 & 10.26 $\pm$ 0.54\\
g (Fig.~\ref{fig:msrg})  && 11.11 $\pm$ 1.22  & 0.26 $\pm$ 0.02  && 0.17 $\pm$ 0.04 & 0.75 $\pm$ 0.27 & 0.07 $\pm$ 0.03 & 8.49 $\pm$ 0.48\\
r (Fig.~\ref{fig:rmall})  && 13.98 $\pm$ 1.73  & 0.25 $\pm$ 0.02  && 0.19 $\pm$ 0.04 & 0.79 $\pm$ 0.30 & 0.05 $\pm$ 0.02 & 13.31 $\pm$ 0.72\\
i (Fig.~\ref{fig:msri})  && 11.69 $\pm$ 1.32  & 0.25 $\pm$ 0.02  && 0.16 $\pm$ 0.05 & 0.69 $\pm$ 0.21 & 0.09 $\pm$ 0.04 & 6.12 $\pm$ 0.26\\
z (Fig.~\ref{fig:msrz})  && 15.36 $\pm$ 1.86  & 0.24 $\pm$ 0.02  && 0.15 $\pm$ 0.04 & 0.71 $\pm$ 0.23 & 0.10 $\pm$ 0.04 & 7.05 $\pm$ 0.29\\
Z (Fig.~\ref{fig:msrzv})  && 24.61 $\pm$ 3.20  & 0.22 $\pm$ 0.02  && 0.16 $\pm$ 0.04 & 0.85 $\pm$ 0.34 & 0.08 $\pm$ 0.03 & 17.47 $\pm$ 0.98\\
Y (Fig.~\ref{fig:msry})  && 19.66 $\pm$ 2.52  & 0.23 $\pm$ 0.02  && 0.16 $\pm$ 0.04 & 0.77 $\pm$ 0.29 & 0.09 $\pm$ 0.03 & 10.97 $\pm$ 0.55\\
J (Fig.~\ref{fig:msrj})  && 19.53 $\pm$ 2.39  & 0.23 $\pm$ 0.02  && 0.15 $\pm$ 0.04 & 0.76 $\pm$ 0.27 & 0.10 $\pm$ 0.04 & 9.32 $\pm$ 0.43\\
H (Fig.~\ref{fig:msrh})  && 15.35 $\pm$ 1.81  & 0.24 $\pm$ 0.02  && 0.15 $\pm$ 0.04 & 0.70 $\pm$ 0.23 & 0.10 $\pm$ 0.04 & 7.12 $\pm$ 0.30\\
K (Fig.~\ref{fig:msrk})  && 10.68 $\pm$ 1.17  & 0.25 $\pm$ 0.02  && 0.14 $\pm$ 0.05 & 0.67 $\pm$ 0.18 & 0.12 $\pm$ 0.06 & 4.72 $\pm$ 0.16\\

	\hline	
\multicolumn{3}{l}{ \textbf{S\'ersic + (\textit{u-r}) colour cut} }\\
u (Fig.~\ref{fig:msru})  && 26.90 $\pm$ 4.21  & 0.22 $\pm$ 0.02  && 0.12 $\pm$ 0.05 & 0.70 $\pm$ 0.22 & 0.24 $\pm$ 0.18 & 6.91 $\pm$ 0.27\\
g (Fig.~\ref{fig:msrg})  && 13.25 $\pm$ 1.43  & 0.24 $\pm$ 0.02  && 0.15 $\pm$ 0.04 & 0.66 $\pm$ 0.17 & 0.11 $\pm$ 0.05 & 5.66 $\pm$ 0.19\\
r (Fig.~\ref{fig:rmall})  && 15.16 $\pm$ 1.71  & 0.24 $\pm$ 0.02  && 0.15 $\pm$ 0.04 & 0.68 $\pm$ 0.19 & 0.11 $\pm$ 0.05 & 6.39 $\pm$ 0.21\\
i (Fig.~\ref{fig:msri})  && 12.93 $\pm$ 1.35  & 0.24 $\pm$ 0.02  && 0.12 $\pm$ 0.05 & 0.66 $\pm$ 0.15 & 0.20 $\pm$ 0.13 & 4.19 $\pm$ 0.12\\
z (Fig.~\ref{fig:msrz})  && 22.86 $\pm$ 2.87  & 0.22 $\pm$ 0.02  && 0.07 $\pm$ 0.04 & 0.66 $\pm$ 0.14 & 0.61 $\pm$ 1.10 & 3.67 $\pm$ 0.09\\
Z (Fig.~\ref{fig:msrzv})  && 25.10 $\pm$ 3.15  & 0.21 $\pm$ 0.02  && 0.11 $\pm$ 0.04 & 0.68 $\pm$ 0.19 & 0.23 $\pm$ 0.15 & 6.16 $\pm$ 0.21\\
Y (Fig.~\ref{fig:msry})  && 21.42 $\pm$ 2.54  & 0.22 $\pm$ 0.02  && 0.11 $\pm$ 0.04 & 0.66 $\pm$ 0.17 & 0.26 $\pm$ 0.19 & 4.97 $\pm$ 0.15\\
J (Fig.~\ref{fig:msrj})  && 19.85 $\pm$ 2.32  & 0.22 $\pm$ 0.02  && 0.08 $\pm$ 0.04 & 0.64 $\pm$ 0.15 & 0.45 $\pm$ 0.54 & 3.62 $\pm$ 0.10\\
H (Fig.~\ref{fig:msrh})  && 17.13 $\pm$ 1.91  & 0.23 $\pm$ 0.02  && 0.09 $\pm$ 0.05 & 0.66 $\pm$ 0.15 & 0.32 $\pm$ 0.29 & 4.08 $\pm$ 0.11\\
K (Fig.~\ref{fig:msrk})  && 13.13 $\pm$ 1.36  & 0.24 $\pm$ 0.02  && 0.09 $\pm$ 0.05 & 0.66 $\pm$ 0.13 & 0.37 $\pm$ 0.37 & 3.42 $\pm$ 0.08\\	
	\hline	
\multicolumn{3}{l}{\textbf{Morphology cut }}\\
u (Fig.~\ref{fig:msru})  && 23.75 $\pm$ 3.29  & 0.23 $\pm$ 0.02  && 0.16 $\pm$ 0.04 & 0.95 $\pm$ 0.33 & 0.11 $\pm$ 0.05 & 19.39 $\pm$ 0.91\\
g (Fig.~\ref{fig:msrg})  && 31.15 $\pm$ 4.11  & 0.21 $\pm$ 0.02  && 0.17 $\pm$ 0.03 & 0.99 $\pm$ 0.38 & 0.08 $\pm$ 0.03 & 32.69 $\pm$ 1.67\\
r (Fig.~\ref{fig:rmall})  && 37.24 $\pm$ 4.82  & 0.20 $\pm$ 0.02  && 0.16 $\pm$ 0.03 & 1.00 $\pm$ 0.37 & 0.10 $\pm$ 0.03 & 33.62 $\pm$ 1.63\\
i (Fig.~\ref{fig:msri})  && 30.10 $\pm$ 3.86  & 0.21 $\pm$ 0.02  && 0.15 $\pm$ 0.04 & 0.96 $\pm$ 0.33 & 0.11 $\pm$ 0.04 & 21.86 $\pm$ 0.98\\
z (Fig.~\ref{fig:msrz})  && 33.46 $\pm$ 4.32  & 0.21 $\pm$ 0.02  && 0.14 $\pm$ 0.04 & 0.95 $\pm$ 0.33 & 0.14 $\pm$ 0.06 & 19.87 $\pm$ 0.86\\
Z (Fig.~\ref{fig:msrzv})  && 66.68 $\pm$ 10.54  & 0.17 $\pm$ 0.02  && 0.14 $\pm$ 0.03 & 0.96 $\pm$ 0.40 & 0.13 $\pm$ 0.04 & 47.94 $\pm$ 2.39\\
Y (Fig.~\ref{fig:msry})  && 48.56 $\pm$ 6.88  & 0.19 $\pm$ 0.02  && 0.15 $\pm$ 0.03 & 0.97 $\pm$ 0.38 & 0.11 $\pm$ 0.04 & 35.10 $\pm$ 1.69\\
J (Fig.~\ref{fig:msrj})  && 41.35 $\pm$ 5.59  & 0.19 $\pm$ 0.02  && 0.13 $\pm$ 0.04 & 0.96 $\pm$ 0.33 & 0.18 $\pm$ 0.09 & 20.02 $\pm$ 0.88\\
H (Fig.~\ref{fig:msrh})  && 31.96 $\pm$ 3.90  & 0.20 $\pm$ 0.02  && 0.14 $\pm$ 0.04 & 0.94 $\pm$ 0.32 & 0.13 $\pm$ 0.06 & 19.18 $\pm$ 0.83\\
K (Fig.~\ref{fig:msrk})  && 20.45 $\pm$ 2.19  & 0.22 $\pm$ 0.02  && 0.14 $\pm$ 0.04 & 0.91 $\pm$ 0.28 & 0.13 $\pm$ 0.05 & 14.03 $\pm$ 0.59\\
	\hline
\end{tabular}
\captionsetup{width=0.5\textwidth}
\caption{The Bayesian expectation parameters for the late-type galaxy \msr relation according to various population definitions. Parameters a and b are used for the single exponential in Eq.~\ref{equ:rm1} and alpha, beta, gamma and $M_0$ for the two component fit in Eq.~\ref{equ:rm2}. Also shown are the parameters found by S03 (S\'ersic cut population only).}
\label{table:rmfitsL}
\end{table*}

\begin{table*}
\centering
\begin{tabular}{|l|l|lll|ll|l|ll|}\\
\\
	\hline
	\multicolumn{3}{l}{\textbf{Early-type galaxies}}&&\\
\\
 Case 	&& a{ }($\mathrm {10}^{-5}$) & b && ${\alpha}$ & ${\beta}$ &$ {\gamma} $& M$_0${ } ($\mathrm  {10}^{10}\mathcal{M_{\sun}}$) \\ 
	\hline \hline
\multicolumn{3}{l}{	\textbf{S\`ersic cut}} \\
u (Fig.~\ref{fig:msru})  && 1345.84 $\pm$ 214.01  & 0.25 $\pm$ 0.03  && 0.07 $\pm$ 0.05 & 0.80 $\pm$ 0.20 & 0.71 $\pm$ 2.19 & 8.43 $\pm$ 0.27\\
g (Fig.~\ref{fig:msrg})  && 8.40 $\pm$ 0.63  & 0.44 $\pm$ 0.02  && 0.10 $\pm$ 0.06 & 0.79 $\pm$ 0.09 & 0.17 $\pm$ 0.13 & 2.54 $\pm$ 0.06\\
r (Fig.~\ref{fig:rmall})  && 8.37 $\pm$ 0.62  & 0.44 $\pm$ 0.02  && 0.10 $\pm$ 0.06 & 0.76 $\pm$ 0.09 & 0.16 $\pm$ 0.12 & 2.42 $\pm$ 0.06\\
i (Fig.~\ref{fig:msri})  && 7.74 $\pm$ 0.53  & 0.44 $\pm$ 0.02  && 0.10 $\pm$ 0.06 & 0.78 $\pm$ 0.09 & 0.18 $\pm$ 0.14 & 2.43 $\pm$ 0.05\\
z (Fig.~\ref{fig:msrz})  && 107.23 $\pm$ 10.03  & 0.34 $\pm$ 0.02  && 0.0003 $\pm$ 0.0002 & 0.84 $\pm$ 0.11 & 2.08 $\pm$ 0.15 & 3.86 $\pm$ 0.07\\
Z (Fig.~\ref{fig:msrzv})  && 16.04 $\pm$ 1.26  & 0.41 $\pm$ 0.02  && 0.10 $\pm$ 0.06 & 0.74 $\pm$ 0.10 & 0.15 $\pm$ 0.11 & 2.85 $\pm$ 0.07\\
Y (Fig.~\ref{fig:msry})  && 11.96 $\pm$ 0.83  & 0.42 $\pm$ 0.02  && 0.11 $\pm$ 0.06 & 0.73 $\pm$ 0.09 & 0.13 $\pm$ 0.09 & 2.48 $\pm$ 0.06\\
J (Fig.~\ref{fig:msrj})  && 27.60 $\pm$ 1.98  & 0.39 $\pm$ 0.02  && 0.09 $\pm$ 0.06 & 0.76 $\pm$ 0.09 & 0.23 $\pm$ 0.20 & 3.07 $\pm$ 0.07\\
H (Fig.~\ref{fig:msrh})  && 36.04 $\pm$ 2.71  & 0.38 $\pm$ 0.02  && 0.08 $\pm$ 0.05 & 0.74 $\pm$ 0.10 & 0.27 $\pm$ 0.24 & 2.99 $\pm$ 0.07\\
K (Fig.~\ref{fig:msrk})  && 23.64 $\pm$ 1.69  & 0.40 $\pm$ 0.02  && 0.10 $\pm$ 0.06 & 0.71 $\pm$ 0.09 & 0.18 $\pm$ 0.13 & 2.56 $\pm$ 0.06\\
		S03 									&& 0.347		& 0.56 			&& - & -& - 	& -\\

	\hline	
	\multicolumn{3}{l}{\textbf{(\textit{u-r}) colour cut}} \\
	
u (Fig.~\ref{fig:msru})  && 7.12 $\pm$ 0.59  & 0.46 $\pm$ 0.03  && 0.12 $\pm$ 0.07 & 0.58 $\pm$ 0.07 & 0.11 $\pm$ 0.09 & 0.98 $\pm$ 0.03\\
g (Fig.~\ref{fig:msrg})  && 5.97 $\pm$ 0.40  & 0.45 $\pm$ 0.02  && 0.10 $\pm$ 0.06 & 0.72 $\pm$ 0.08 & 0.17 $\pm$ 0.13 & 1.94 $\pm$ 0.05\\
r (Fig.~\ref{fig:rmall})  && 7.32 $\pm$ 0.50  & 0.44 $\pm$ 0.02  && 0.10 $\pm$ 0.06 & 0.75 $\pm$ 0.09 & 0.17 $\pm$ 0.13 & 2.31 $\pm$ 0.05\\
i (Fig.~\ref{fig:msri})  && 4.75 $\pm$ 0.32  & 0.46 $\pm$ 0.02  && 0.09 $\pm$ 0.06 & 0.77 $\pm$ 0.08 & 0.18 $\pm$ 0.13 & 2.17 $\pm$ 0.05\\
z (Fig.~\ref{fig:msrz})  && 7.34 $\pm$ 0.52  & 0.44 $\pm$ 0.02  && 0.09 $\pm$ 0.06 & 0.76 $\pm$ 0.09 & 0.19 $\pm$ 0.15 & 2.36 $\pm$ 0.05\\
Z (Fig.~\ref{fig:msrzv})  && 11.98 $\pm$ 0.89  & 0.42 $\pm$ 0.02  && 0.10 $\pm$ 0.06 & 0.66 $\pm$ 0.08 & 0.15 $\pm$ 0.11 & 1.94 $\pm$ 0.05\\
Y (Fig.~\ref{fig:msry})  && 8.47 $\pm$ 0.58  & 0.43 $\pm$ 0.02  && 0.10 $\pm$ 0.06 & 0.70 $\pm$ 0.08 & 0.16 $\pm$ 0.11 & 2.09 $\pm$ 0.05\\
J (Fig.~\ref{fig:msrj})  && 6.62 $\pm$ 0.46  & 0.44 $\pm$ 0.02  && 0.10 $\pm$ 0.06 & 0.69 $\pm$ 0.08 & 0.14 $\pm$ 0.09 & 1.80 $\pm$ 0.04\\
H (Fig.~\ref{fig:msrh})  && 7.62 $\pm$ 0.52  & 0.44 $\pm$ 0.02  && 0.11 $\pm$ 0.06 & 0.64 $\pm$ 0.07 & 0.12 $\pm$ 0.07 & 1.55 $\pm$ 0.04\\
K (Fig.~\ref{fig:msrk})  && 4.83 $\pm$ 0.33  & 0.46 $\pm$ 0.02  && 0.10 $\pm$ 0.06 & 0.68 $\pm$ 0.07 & 0.13 $\pm$ 0.09 & 1.58 $\pm$ 0.04	\\
		\hline	
\multicolumn{3}{l}{	\textbf{(\textit{g-i}) colour cut} }\\

u  (Fig.~\ref{fig:msru})  && 10.03 $\pm$ 0.86  & 0.44 $\pm$ 0.03  && 0.12 $\pm$ 0.07 & 0.59 $\pm$ 0.07 & 0.13 $\pm$ 0.10 & 1.16 $\pm$ 0.03\\
g  (Fig.~\ref{fig:msrg})  && 7.46 $\pm$ 0.51  & 0.44 $\pm$ 0.02  && 0.10 $\pm$ 0.06 & 0.73 $\pm$ 0.08 & 0.18 $\pm$ 0.13 & 2.09 $\pm$ 0.05\\
r  (Fig.~\ref{fig:rmall})  && 8.25 $\pm$ 0.57  & 0.44 $\pm$ 0.02  && 0.10 $\pm$ 0.06 & 0.74 $\pm$ 0.08 & 0.17 $\pm$ 0.13 & 2.24 $\pm$ 0.05\\
i  (Fig.~\ref{fig:msri})  && 5.40 $\pm$ 0.35  & 0.46 $\pm$ 0.02  && 0.10 $\pm$ 0.06 & 0.78 $\pm$ 0.08 & 0.17 $\pm$ 0.13 & 2.30 $\pm$ 0.05\\
z  (Fig.~\ref{fig:msrz})  && 8.79 $\pm$ 0.62  & 0.44 $\pm$ 0.02  && 0.09 $\pm$ 0.06 & 0.77 $\pm$ 0.09 & 0.21 $\pm$ 0.17 & 2.48 $\pm$ 0.05\\
Z  (Fig.~\ref{fig:msrzv})  && 13.16 $\pm$ 0.97  & 0.42 $\pm$ 0.02  && 0.10 $\pm$ 0.06 & 0.68 $\pm$ 0.08 & 0.14 $\pm$ 0.10 & 2.15 $\pm$ 0.05\\
Y  (Fig.~\ref{fig:msry})  && 9.95 $\pm$ 0.71  & 0.43 $\pm$ 0.02  && 0.10 $\pm$ 0.06 & 0.68 $\pm$ 0.08 & 0.14 $\pm$ 0.10 & 1.92 $\pm$ 0.05\\
J  (Fig.~\ref{fig:msrj})  && 7.50 $\pm$ 0.53  & 0.44 $\pm$ 0.02  && 0.09 $\pm$ 0.06 & 0.73 $\pm$ 0.08 & 0.16 $\pm$ 0.11 & 2.17 $\pm$ 0.05\\
H  (Fig.~\ref{fig:msrh})  && 8.61 $\pm$ 0.60  & 0.43 $\pm$ 0.02  && 0.10 $\pm$ 0.06 & 0.65 $\pm$ 0.07 & 0.13 $\pm$ 0.09 & 1.66 $\pm$ 0.04\\
K  (Fig.~\ref{fig:msrk})   && 5.39 $\pm$ 0.36  & 0.45 $\pm$ 0.02  && 0.10 $\pm$ 0.06 & 0.71 $\pm$ 0.08 & 0.14 $\pm$ 0.09 & 1.79 $\pm$ 0.04\\
	\hline	
\multicolumn{3}{l}{\textbf{S\'ersic + (\textit{u-r}) colour cut} }\\
u (Fig.~\ref{fig:msru})  && 24.46 $\pm$ 2.87  & 0.41 $\pm$ 0.03  && 0.11 $\pm$ 0.07 & 0.61 $\pm$ 0.11 & 0.19 $\pm$ 0.19 & 2.39 $\pm$ 0.08\\
g (Fig.~\ref{fig:msrg})  && 0.23 $\pm$ 0.02  & 0.58 $\pm$ 0.03  && 0.12 $\pm$ 0.08 & 0.76 $\pm$ 0.07 & 0.08 $\pm$ 0.05 & 1.33 $\pm$ 0.04\\
r (Fig.~\ref{fig:rmall})  && 0.30 $\pm$ 0.02  & 0.57 $\pm$ 0.03  && 0.12 $\pm$ 0.07 & 0.78 $\pm$ 0.08 & 0.08 $\pm$ 0.05 & 1.54 $\pm$ 0.04\\
i (Fig.~\ref{fig:msri})  && 0.22 $\pm$ 0.01  & 0.58 $\pm$ 0.03  && 0.12 $\pm$ 0.07 & 0.78 $\pm$ 0.08 & 0.08 $\pm$ 0.05 & 1.46 $\pm$ 0.04\\
z (Fig.~\ref{fig:msrz})  && 0.39 $\pm$ 0.03  & 0.56 $\pm$ 0.03  && 0.12 $\pm$ 0.08 & 0.74 $\pm$ 0.08 & 0.08 $\pm$ 0.06 & 1.43 $\pm$ 0.04\\
Z (Fig.~\ref{fig:msrzv})  && 0.56 $\pm$ 0.04  & 0.54 $\pm$ 0.03  && 0.12 $\pm$ 0.08 & 0.70 $\pm$ 0.08 & 0.07 $\pm$ 0.04 & 1.33 $\pm$ 0.04\\
Y (Fig.~\ref{fig:msry})  && 0.42 $\pm$ 0.03  & 0.55 $\pm$ 0.03  && 0.12 $\pm$ 0.08 & 0.70 $\pm$ 0.07 & 0.07 $\pm$ 0.04 & 1.16 $\pm$ 0.03\\
J (Fig.~\ref{fig:msrj})  && 0.50 $\pm$ 0.04  & 0.55 $\pm$ 0.03  && 0.12 $\pm$ 0.08 & 0.72 $\pm$ 0.08 & 0.07 $\pm$ 0.05 & 1.42 $\pm$ 0.04\\
H (Fig.~\ref{fig:msrh})  && 0.51 $\pm$ 0.04  & 0.55 $\pm$ 0.03  && 0.12 $\pm$ 0.08 & 0.68 $\pm$ 0.07 & 0.06 $\pm$ 0.04 & 1.18 $\pm$ 0.04\\
K (Fig.~\ref{fig:msrk})  && 0.36 $\pm$ 0.03  & 0.56 $\pm$ 0.03  && 0.13 $\pm$ 0.09 & 0.68 $\pm$ 0.07 & 0.05 $\pm$ 0.03 & 0.96 $\pm$ 0.03\\
	\hline	
\multicolumn{3}{l}{\textbf{Morphology cut} }\\
u (Fig.~\ref{fig:msru})  && 4.84 $\pm$ 0.40  & 0.47 $\pm$ 0.03  && 0.12 $\pm$ 0.07 & 0.69 $\pm$ 0.09 & 0.11 $\pm$ 0.08 & 1.70 $\pm$ 0.05\\
g (Fig.~\ref{fig:msrg})  && 3.77 $\pm$ 0.25  & 0.47 $\pm$ 0.02  && 0.11 $\pm$ 0.07 & 0.76 $\pm$ 0.09 & 0.11 $\pm$ 0.07 & 2.01 $\pm$ 0.05\\
r (Fig.~\ref{fig:rmall})  && 4.19 $\pm$ 0.28  & 0.46 $\pm$ 0.02  && 0.11 $\pm$ 0.06 & 0.78 $\pm$ 0.09 & 0.12 $\pm$ 0.08 & 2.25 $\pm$ 0.06\\
i (Fig.~\ref{fig:msri})  && 2.44 $\pm$ 0.15  & 0.49 $\pm$ 0.02  && 0.11 $\pm$ 0.06 & 0.79 $\pm$ 0.08 & 0.12 $\pm$ 0.08 & 1.96 $\pm$ 0.05\\
z (Fig.~\ref{fig:msrz})  && 4.54 $\pm$ 0.31  & 0.46 $\pm$ 0.02  && 0.11 $\pm$ 0.06 & 0.76 $\pm$ 0.09 & 0.12 $\pm$ 0.08 & 2.13 $\pm$ 0.05\\
Z (Fig.~\ref{fig:msrzv})  && 6.74 $\pm$ 0.50  & 0.44 $\pm$ 0.03  && 0.11 $\pm$ 0.06 & 0.75 $\pm$ 0.10 & 0.11 $\pm$ 0.07 & 2.30 $\pm$ 0.06\\
Y (Fig.~\ref{fig:msry})  && 4.97 $\pm$ 0.36  & 0.45 $\pm$ 0.02  && 0.11 $\pm$ 0.06 & 0.73 $\pm$ 0.09 & 0.10 $\pm$ 0.06 & 2.03 $\pm$ 0.05\\
J (Fig.~\ref{fig:msrj})  && 3.73 $\pm$ 0.24  & 0.46 $\pm$ 0.02  && 0.11 $\pm$ 0.06 & 0.74 $\pm$ 0.09 & 0.10 $\pm$ 0.06 & 1.89 $\pm$ 0.05\\
H (Fig.~\ref{fig:msrh})  && 4.08 $\pm$ 0.28  & 0.46 $\pm$ 0.02  && 0.11 $\pm$ 0.07 & 0.71 $\pm$ 0.08 & 0.09 $\pm$ 0.05 & 1.79 $\pm$ 0.05\\
K (Fig.~\ref{fig:msrk})  && 2.64 $\pm$ 0.18  & 0.48 $\pm$ 0.02  && 0.11 $\pm$ 0.06 & 0.72 $\pm$ 0.08 & 0.09 $\pm$ 0.05 & 1.57 $\pm$ 0.04\\
	\hline
\end{tabular}
\captionsetup{width=0.5\textwidth}
\caption{The Bayesian expectation parameters for the early-type galaxy \msr relation according to various population definitions. Parameters a and b are used for the single exponential in Eq.~\ref{equ:rm1} and alpha, beta, gamma and $M_0$ for the two component fit in Eq.~\ref{equ:rm2}. Also shown are the parameters found by S03 (S\'ersic cut population only).}
\label{table:rmfitsE}
\end{table*}

To  be consistent with the previous work of S03 the separating S\'ersic index was set to
n=2.5, the average of the exponential profile (n=1) and de Vaucouleurs
profile (n=4). In the \textit{r}-band this splits our sample into 6108 late-type galaxies and 
 2291  early-type galaxies.

Fig. \ref{fig:select} (upper panel) shows the stellar mass versus
S\'ersic index distribution of our sample, colour coded by the dust corrected restframe
(\textit{u-r})$_{stars}$ colour, the blue dashed line indicates the chosen
S\'ersic separator (n=2.5). The (\textit{u-r})$_{stars}$ restframe colour was taken from \citet{Taylor2011}.\\

 Fig. \ref{fig:rmall}a shows the resulting
\msr relations, where the left panel shows the late-type galaxies with n$<$2.5 in
blue and the right panel shows the early-type galaxies with n$>$2.5 in red.\footnote{We 
caution again that using the S\'ersic index to establish early- and late-type galaxy populations is misleading when assuming morphological agreement since there are elliptical galaxies with low n and disc galaxies with high n.} 
The individual galaxies are plotted as small dots, the coloured squares show median 
binned data for visualisation only (the fitting is performed on the entire sample)
 with the dispersion of the data
shown as black error bars representing the 0.25 and 0.75 quantile.
The error on the median data points is shown as orange error bars
(often smaller than the data point). The contours show the weighted
90$^{th}$, 68$^{th}$ and 50$^{th}$ percentile of the highest density region (HDR) of the data. The best
fit lines (via Bayesian parameter expectation) to the data are shown in red 
and blue for Eq.~\ref{equ:rm1} (single power law)
and in green for Eq.~\ref{equ:rm2} (two component power law). 
The black lines show the \msr relation as found by S03 (the dot-dashed line is corrected for size difference and the solid line is uncorrected, see below for explanation). 
The grey dashed vertical line on the plot indicates the lower mass limit
which was calculated for an unbiased volume limited sample out to z=0.1. We
plot this line to visually illustrate the point at which the
volume becomes reduced, but remind the reader that we fit to the entire mass-range of the staggered volume limited sample shown. The fitting parameters to
Eqs. \ref{equ:rm1} and \ref{equ:rm2} can be found in Table
\ref{table:rmfitsL} for our late-type sample and Table \ref{table:rmfitsE} for the early-type sample. For comparison both Tables also show the respective early- and late-type \msr fitting parameters found by S03.\\

It is important to note that the S03 \msr relation was fitted using the \textit{z}-band circularised half-light radius thus a direct comparison between the \msr relation presented in Fig. \ref{fig:rmall}a and S03 would lead to wrong conclusions since we expect the \textit{z}-band sizes to be smaller than the \textit{r}-band sizes.  
 
To illustrate the difference introduced by analysing the \msr relation in different wave bands we have plotted the S03 \msr relation without any correction of the expected sizes (i.e. \textit{z}-band sizes; black solid line, Fig.~\ref{fig:rmall}a) and with sizes corrected to reflect \textit{r}-band sizes (black dot-dashed line, Fig.~\ref{fig:rmall}a).
To correct the S03 \msr relation we use the wavelength dependent size relation for discs and spheroids found by \citet{Kelvin2012} to establish a ratio of the sizes between
the r- and \textit{z}-band of 1.075 for the late-types and 1.123 for the early-types sizes. We then multiply the sizes obtained for the S03 late-type relation by these ratios.
The resulting shift moves the early-type S03 \msr relation further onto our galaxy distribution, however, it is steeper than our observed relation. For late types we still see an offset between the S03 \msr relation and our data. 

\begin{figure}
\centering
\includegraphics[width=0.49\textwidth]{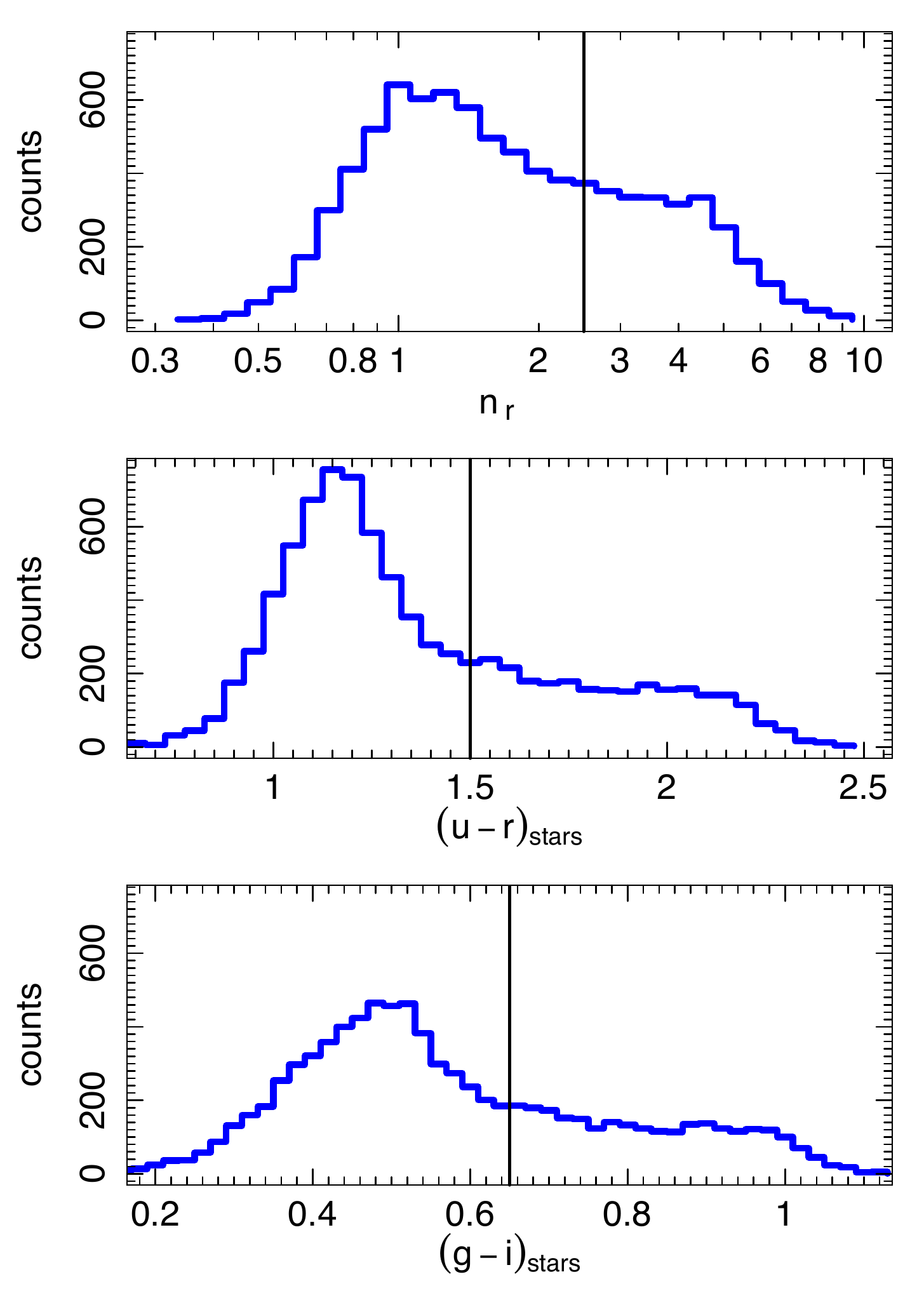}
\caption{The top panel shows the histogram for the S\'ersic index distribution in the \textit{r}-band. 
There is no clear bimodality visible in the distribution, i.e.~there is no trough between the two populations. Instead number counts plateau after the initial peak (n$\sim$1) before falling off further after the second `peak' (n$\sim$4). The black vertical line shows the chosen S\'ersic index separator n=2.5.\newline 
The middle panel shows the histogram of
the (\textit{u-r})$_{stars}$ dust corrected restframe colour distribution. 
As before there is no clear bimodality, however, the first peak seems clearer than in the S\'ersic index distribution. The vertical black line indicates the used threshold of 1.5. \newline
The bottom panel shows the histogram of
the (\textit{g-i})$_{stars}$ dust corrected restframe colour distribution. Again the bimodality is not very clear and the threshold set to 0.65 (black vertical line).}
\label{fig:nhist}
\end{figure}

Fig.~\ref{fig:msrz} (top panel) shows the direct comparison between the S03 and our \msr relation in the \textit{z}-band. Even though the same waveband is compared here we still see an offset between the two relations. For the early-types this equates to S03 sizes being on average ~1.1kpc smaller than our sizes at most galaxy masses but larger at $\mathcal{M_*}\gtrsim 2\times10^{11}\mathcal{M_{\sun}}$. However, in this regime our \msr relation is not well constrained. 
For late-type galaxies we have a median size offset between S03 and our sizes of $\sim$0.9kpc.
The main contributing factors to this discrepancy are likely to be our deeper data and the use of elliptical semi-major axis R$_e$ as opposed to circularised R$_e$ used in S03. The former causes the observed differences in the slope while the latter shifts our \msr relation to larger sizes. Using elliptical semi-major axis radii instead of circularised sizes also explains the larger size offset for late-type galaxies, which on average have a higher (observed) ellipticity than the early-type galaxies.
Also note that for a fair comparison, the fits should only be considered in the mass range in which the S03 relation was established, these are  $\log_{10}(\mathcal{M}_*\slash{}\mathcal{M}_{\sun})\gtrsim8.8$  for late-types and $\log_{10}(\mathcal{M}_*\slash\mathcal{M}_{\sun})\gtrsim10.1$ for early-types.\\

For the early-types, S03 found a single power law (Eq.~\ref{equ:rm1}) to be a good fit. In Fig.~\ref{fig:rmall}, if we consider the same mass range then the S03 \msr relation seemingly fits well onto our data. However, if we consider the entire mass range available we find that the two component power law (Eq.~\ref{equ:rm2}) is a better fit due to some flattening in the \msr distribution observed for low mass galaxies (in particular galaxies below $\log_{10}(\mathcal{M}_*\slash\mathcal{M}_{\sun})\lesssim10)$ and a steepening of the relation at the high-mass end. A similar flattening was also observed for spheroids by e.g. \cite{Shankar2013} and \cite{Berg2014} and could be related to dissipation processes during (gas-rich) mergers. In fact, it has been known for some time that the elliptical \msr relation becomes flat for small galaxies, especially when considering dwarf ellipticals (see section \ref{sec:flat} for more information).
 However, small elliptical galaxies ($\mathcal{M}<10^{10}\mathcal{M}_{\sun}$) have been found to also have smaller S\'ersic indices (n$<$2.5) (see e.g.~\citealt{Graham2006})
  and thus the flattening seen here is likely caused by cross-scattering of not-elliptical galaxies with higher S\'ersic indices. In addition, the flattening observed in our sample is based on very few galaxies
which cannot constrain the \msr relation fit to Eq.~\ref{equ:rm2} well and cause the fit to Eq.~\ref{equ:rm1} to flatten considerably.   

For the late-types, when considering the same mass-range, our data shows a similar distribution (see Fig.~\ref{fig:rmall}) to S03, who found a two component power law to be a good fit to the data. However,  we find that our data, at fixed mass, has larger sizes than the S03 relation, even after the correction for the wavelength dependent sizes (due to the use of circularized sizes in the S03 fit). 
In addition, the fit to Eq.~\ref{equ:rm2} has a high `transitioning mass' M$_0$, of the order of a few $10^{11}\mathcal{M}_{\sun}$, which lies beyond most galaxy in our sample (at least 99.5\% of galaxies have masses below M$_0$ in any band). Hence, over the mass range observed, the two component fit (Eq.~\ref{equ:rm2}) is driven to a single exponential fit with slope $\alpha$ in our MCMC fitting. 
In addition, fitting parameter $\alpha$ is, within the errors, not dissimilar to fitting parameter b from Eq.~\ref{equ:rm1}.
This makes the fit to the two-component power law superfluous over the mass range observed here.

\subsubsection{Alternative S\'ersic population separators}

As pointed out previously, the flattening observed in the early-type \msr relation in our S\'ersic index divided sample might be due to the inclusion of galaxies that in reality belong to the morphologically classified late-type population.
One possible cause of this is that a separation of the population at n=2.5 is a poor description of the actual distribution of the S\'ersic indices in our sample. We expect a bimodality in the S\'ersic index distribution with late-type galaxies tending to n=1 and early-type galaxies to n=4. To check this we plot the S\'ersic index distribution in the top panel of Fig.~\ref{fig:nhist}. However, we see no clear separating S\'ersic index between early- and late-type populations.

Bimodalities are most evident when the two populations are seen in equal numbers. 
However, over the whole mass range probed in our sample we have more late-type than early-type galaxies especially at lower masses. In addition, elliptical galaxies tend to have lower S\'ersic indices at lower masses ( $\mathcal{M}\lesssim10^{10}\mathcal{M}_{\sun}$). Hence, including these low-mass galaxies will skew the distribution of S\'ersic indices towards smaller numbers making the bimodality less obvious. This can also be seen in the top panel of Fig.~\ref{fig:select}, which shows that galaxies with high S\'ersic indices (n$>$2.5) tend to have masses above $10^{10}\mathcal{M}_{\sun}$ and most galaxies with masses $<10^9\mathcal{M}_{\sun}$ have S\'ersic indices n$<$2.5. The few galaxies that have low masses ($\sim 10^{10} \rm \mathcal{M}_{\sun}$) and high S\'ersic indices (n$>$2.5) are the `cross-scatter' we see in the above \msr relation.

Fig.~\ref{fig:eyeball} shows the same data as Fig.~\ref{fig:select}, however, the data points are colour coded by the visual classification. The top panel shows that there is a lot of cross-scatter of morphological late-types (i.e.~not-elliptical galaxies) into the high-n region as well as morphological early-types (i.e.~elliptical galaxies) into the low-n region.

Since the dispersion around the mean for late-types is already large, including these cross-scattered galaxies in the late-type sample has little effect. 
However, including the cross-scattered galaxies in the early-type sample increases the dispersion and changes the \msr relation, especially at lower masses.
Considering this we find that an alternative (but rigid) S\'ersic index cut would not improve the \msr relation fits and we will concentrate on other possible population separators which are discussed in the following sections.

\subsection{\msr Relation: division by colour }
\label{sec:colour}

Here we investigate the identification of early- and late-type galaxies depending on two different colour selections. 
We have adopted the dust corrected (\textit{u-r})$_{stars}$ colour division and the dust corrected (\textit{g-i})$_{stars}$ colour division.\\

The two middle panels of Fig.~\ref{fig:select} show the 3D distribution
of the dust corrected (\textit{u-r})$_{stars}$ colour and (\textit{g-i})$_{stars}$ 
colour vs galaxy stellar mass with the data points coloured 
by their S\'ersic index (panel b and c respectively). Both colour distributions show that, in comparison to the S\'ersic index distribution, a distinction between late- and early-types should be clearer with two unconnected population centres visible in the plot.
The middle and lower panel of Fig. \ref{fig:nhist} show the histograms of the (\textit{u-r})$_{stars}$ and (\textit{g-i})$_{stars}$ colours respectively. The peak of the late-type population appears somewhat clearer in the colour histograms than it is in the S\'ersic index histogram and we chose the population division at a point were the late-type populations become reduced and starts to plateau towards the early-type population. 
We set the population cuts to (\textit{u-r})$_{rest}=1.5$ and (\textit{g-i})$_{stars}$=0.65.

This population separation results in 5912 late-type galaxies and 2487 early-type galaxies using the (\textit{u-r})$_{stars}$ colour division and 5876 late-types and 2523 early-types using the (\textit{g-i})$_{stars}$ colour division. 
The \msr fit to the early- and late types divided by colour can be seen in Fig.~\ref{fig:rmall}b for the (\textit{u-r})$_{stars}$ colour cut and panel c for the (\textit{g-i})$_{stars}$ colour cut. The fit parameters are given in Tables \ref{table:rmfitsL} and \ref{table:rmfitsE}, we fit the same equations as for the S\'ersic division (Eq.~\ref{equ:rm1} and \ref{equ:rm2}).\\

Comparing the \msr relations derived using a colour division to those derived by a S\'ersic division we find a reduced number of galaxies at low masses ($\mathcal{M_*}\gtrsim10^{9.4}\mathcal{M_{\sun}}$, as these galaxies have been moved into the late-type sample. 
However, these additions to the late-type sample lead to a slight steepening in the \msr relation for the single exponential fit and the transition mass `M0' in the double power law fit is reduced to the order of several $10^{10}\mathcal{M}_{\sun}$ which is at the upper limit of our data. Overall the fit to Eq.~\ref{equ:rm1} is a good approximation of the \msr relation for late-types, especially in the lower mass range (when compared to the curved relation).
The early-type \msr relations of both colour cuts continue to show some flattening for galaxies with $\mathcal{M_*}\gtrsim2\times10^{10}\mathcal{M_{\sun}}$ and the double power-law fits remain largely unchanged compared to the S\'ersic index early-types. However, for most bands we observe a steepening of the single power law fit to the data. This is largely due to the move of low-mass ($\mathcal{M}\lesssim 10^{9.4}\mathcal{M}_{\sun}$) galaxies into the late-type sample. 
Overall the single power law is a good approximation of the data. However, due to the low-mass flattening of the \msr distribution the single power law fit is not steep enough to fit very massive galaxies (with $\mathcal{M}> 10^{11}\mathcal{M}_{\sun}$) well and hence the double power law fit should be considered instead.

\newpage
\subsection{\msr Relation: Combined S\'ersic index and colour division}
\label{sec:Roll}

A rigid cut by either colour or S\'ersic index
will never be a good representation for early- and late-type galaxy
populations, especially since the early-/late-type classification itself is not rigid.
Figures \ref{fig:select} and \ref{fig:nhist} show that neither the S\'ersic index nor the colour are definitive separators for the early- and late-type populations. The S\'ersic index in particular does not show a clear bimodality and the colour distributions show a slightly sharper peak for the blue galaxies which plateaus and then transitions into the red galaxies.
This is not surprising if we take into account that often early-types are associated with elliptical galaxies and late types with non-elliptical galaxies \citep{Robotham2013}, this will lead to a significant overlap of the populations if only colour or S\'ersic index are  considered as a true representation of the galaxy morphology. 
This point is discussed in detail by \citep{Taylor2014}
and can be seen in the \textit{r}-band S\'ersic index vs $(\textit{u-r})_{rest}$ colour
plot (bottom panel of Fig.~\ref{fig:select}). The plot shows two populations,
one in the blue colour and low S\'ersic index region and the other in the red colour
and high S\'ersic index region. In the plot the data points are
coloured according to their mass also showing that most early-types (i.e.~high-n and red)
are more massive than late-type (i.e.~low-n and blue) galaxies. The contours show the data
density and the blue dashed lines show the previously chosen separators for S\'ersic index and colour.
The plot shows that choosing the (\textit{u-r})$_{stars}$ colour as a separator
reduces the cross-contamination compared to the
S\'ersic index cut. But it is also clear that neither colour nor S\'ersic index are ideal separators and a combined S\'ersic index and colour cut should improve the separation of the early- and late-types.
The solid black line in the bottom panel of Fig.~\ref{fig:select} shows a separation of
the two populations that depends on both the (\textit{u-r})$_{stars}$ colour and
the S\'ersic index. 
It is a `best population division' line, with a slope that is orthogonal to the connecting line
between the two population centres (marked by the crosses) and an intercept that is chosen in such a way that the bijective assignment to the (visually classified) morphological elliptical/ not-elliptical classification (see section \ref{sec:morph}) is maximised, i.e. the probability of correctly assigning a galaxy as either early- or late-type is maximised. 
The resulting division line splits the sample into 6748 late-type
galaxies and 1651 early-type galaxies.
This division line is calculated for each band and the equation is given in panel d) on all \msr relation plots.\\

The resulting \msr relation is plotted in Fig.~\ref{fig:rmall}d. 
There are even less low-mass galaxies included in the early-type population compared to previous cuts. This leads to a steepening of the fit to  Eq.~\ref{equ:rm1}, whereas the fitting parameters to Eq.~\ref{equ:rm2} continue to remain mostly unchanged. 
The fitting parameters to the single power law for the late-types also remain largely unchanged. Whereas the double component power law still has a fairly high `transitioning mass' $M_0$ ($\sim10^{10}\mathcal{M_{\sun}}$) and a slope $\alpha$ that us to shallow to describe the low mass galaxies well ($\mathcal{M}\lesssim10^9\mathcal{M_{\sun}}$).
Using the rolling S\'ersic index and colour cut we find that a single power law fit to the data is sufficient to describe the \msr distribution of both the early- and late-types. 

Comparing the fitting parameters for all the above discussed cases shows that they are quite robust to changes in the chosen population separator, that is if we consider Eq.~\ref{equ:rm1} for late-types and  Eq.~\ref{equ:rm2} for early-types only. The more dominant changes come from the chosen sample, e.g.~the mass range probed or circular vs semi-major axis radii. This becomes apparent when comparing our sample with the S03 relation. For example if we compare the single power law fit for the early-type galaxies in this section with the fit found by S03 we find that the slope is comparable due to the exclusion of many low-mass galaxies in this particular sample selection. \\

 The remaining question is, are any of the chosen separators in fact good enough to describe
the underlying populations satisfactorily, i.e. how do the above
\msr relations compare to the relations found for a visually classified morphological early-/late-type sample? 

\subsection{\msr Relation: division by morphology}
\label{sec:morph}

  \begin{figure}
\centering
\includegraphics[width=0.45\textwidth]{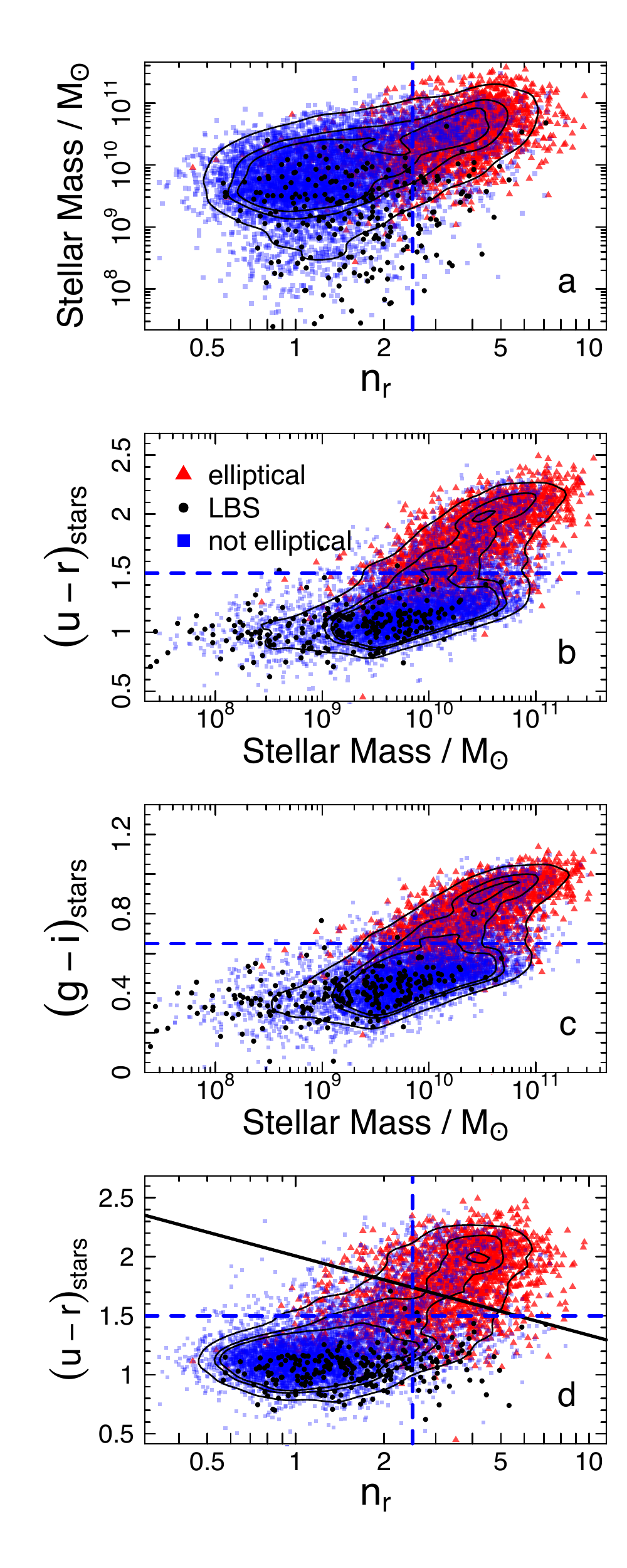}
\caption{The distribution of the galaxies in our sample plotted for the same
sub-plots as Fig.~\ref{fig:select} but with the data points colour coded
according to the visual classification assigned to them. The blue dashed lines indicate 
the chosen S\'ersic index colour population separators and
the black solid line in the lower panel shows the combined (\textit{u-r})$_{stars}$ 
colour and S\'ersic index cut. 
\newline  Two distinct populations of elliptical and `not-elliptical' galaxies can be seen.
In addition we see a population of little blue spheroids which mostly scatter 
across the `not-elliptical' population but in the case of a S\'ersic cut also 
scatter onto the elliptical population.}
\label{fig:eyeball}
\end{figure}

We use the elliptical\slash{}not-elliptical visual classifications as
defined by \cite{Driver2013} who used H\textit{ig} colour images to
classify the GAMAmid sample.  Our morphological sample consists of 2010 elliptical
galaxies, 6151 not-ellipticals and 231 little blue spheroids (LBS hereafter). 
LBS are galaxies that look spheroidal (i.e. elliptical-like) but are blue in colour and typically small (median size $\sim$1.3kpc) and do not fit in well with either our elliptical or not-elliptical sample (see \citealt{Kelvin2014} for initial identification of this sample in GAMA and Moffettt et.~al in prep.~for more details on the nature of these galaxies).

Fig.~\ref{fig:eyeball} shows the population 
distribution in four different panels (as Fig.~\ref{fig:diagnostics}), from top to bottom these are: 
stellar mass vs S\'ersic index, (\textit{u-r})$_{stars}$ colour vs stellar mass, 
(\textit{g-i})$_{stars}$ colour vs stellar mass and
(\textit{u-r})$_{stars}$ colour vs S\'ersic index. 
The galaxies classified as ellipticals are shown in red, not-elliptical in blue and the LBS are black. A significant cross scatter of the elliptical and not-elliptical galaxies can be seen in all plots. 
This means that around 30 to 40\% of the galaxies classified as early-types using a rigid population separator are actually 'not-elliptical' galaxies according to their visual classification. 

However, even though the size of the cross-scatter is similar in all three cases the S\'ersic index cut has the worst sample contamination due to the number of LBS galaxies and other low-mass but high-n not-elliptical galaxies classified as early-type.  
The inclusion of the LBS and low-mass but high-n galaxies influences the early-type fit which can be seen in the S\'ersic index cut \msr relation as the low mass flattening discussed previously. 
The presence of low-n and low-mass elliptical galaxies we see is also expected, see e.g. \cite{Graham2003} who show that there is a continuous downward trend of the S\'ersic index with luminosity (their Fig. 10). However, their inclusion in the late-type sample and exclusion from the early-type sample is not a driving factor in the \msr relation fit of the late-types.

In the case of the colour cuts and the rolling colour and S\'ersic index cut, the late-type galaxies misclassified as early-types are not as clearly distinguishable from the actual elliptical galaxies, i.e. there are less outliers like red and low-mass or red and low-n galaxies. The distribution of these misclassified `early-types' in the stellar mass -- colour space and the colour -- S\'ersic index space is similar to that of the ellipticals, hence the resulting \msr relations have less low mass contamination.\\

 We fit the \msr relation according to the visual
classification and the resulting fits can be seen in
Fig.~\ref{fig:rmall}e and the fitting parameters are given in Tables
\ref{table:rmfitsL} and \ref{table:rmfitsE}. 

The \msr relation fit to the early- and late-type populations according to their
visual classifications shows that early-type galaxies have a
distribution with little scatter but late-type galaxies still
display a large dispersion. 
As seen in the  previous sections the fitting parameters remain relatively robust to the 
slight changes in the overall population sample. 
In addition, the double power law fit to the late-types has a very high value for M$_0$ (a few $10^{11}\mathcal{M}_{\sun}$) which means the fit tends to a single power-law over the mass range observed.
We do again observe the turn off and flattening of the early-type relation, hence we recommend using the double-power law fit to the early-type galaxies. If however a single power law fit is required for comparisons we caution that the relation shown here underestimates the very high mass end of the distribution. If these galaxies are of particular interest we provide a single power law fit to the early-type (late-type) \msr relation for galaxies with $\mathcal{M_*}>2\times10^{10}\mathcal{M}_{\sun}$ ($\mathcal{M_*}>2.5\times10^{9}\mathcal{M}_{\sun}$) in Appendix \ref{app:msr_additional}.

\subsubsection{The low mass flattening of the elliptical \msr relation}
\label{sec:flat}

Fig.~\ref{fig:rmall} shows that early-type galaxies (right hand panel) show evidence of a low mass ($\mathcal{M_*}\lesssim10^{10}\mathcal{M}_{\sun}$) turn off in the \msr relation. Hence a curved relation fit is needed when lower mass early-type galaxies are present in the sample. However, we caution again that not all early-type descriptors represent the underlying elliptical population well and the low mass end of the distribution should be treated with care.\\

Here we show that elliptical galaxies indeed have a flattened \msr relation at the low mass end ($\mathcal{M_*}<10^{10}\mathcal{M}_{\sun}$, see e.g.~\citealt{Graham2013} and references therein) and that the apparent turn off visible in our distribution of early-type galaxies is in good agreement with a sample of low mass elliptical galaxies from \cite{Graham2006}.

\begin{figure}
\includegraphics[width=0.49\textwidth]{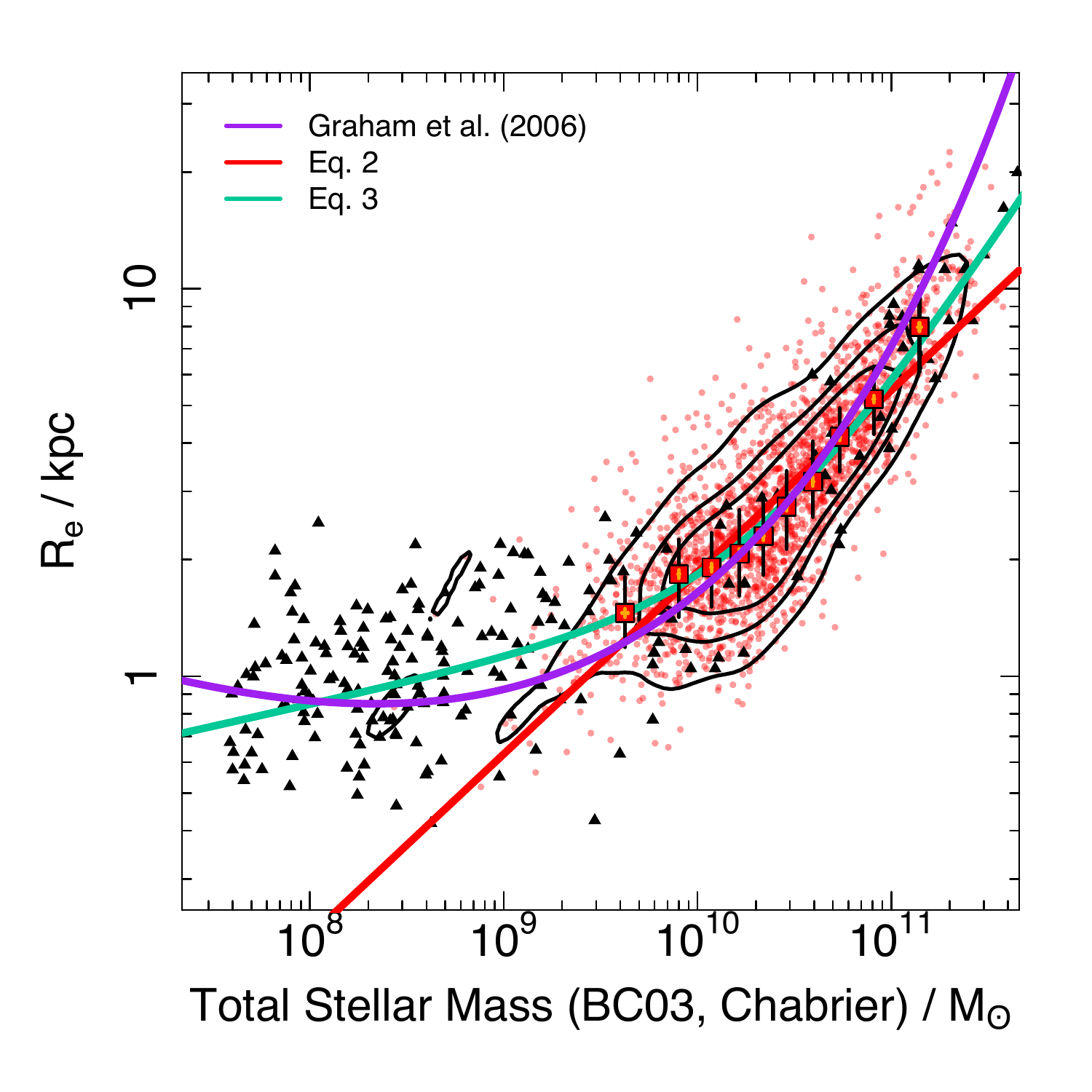}
\caption{The plot shows a comparison of our data (same as bottom panel in Fig.~\ref{fig:msrg}) with data from \citet{Graham2006} shown as black triangles with their predicted curved relation in purple. 
For the mass-range probed in this paper the turn off is not very prominent and a linear \msr relation is a good approximation for the \msr distribution. However, if more low mass ellipticals are included then the flattening of the \msr relation becomes evident and a curved relation is needed to fit the data well.}
\label{fig:curve_compare}
\end{figure}

Fig.~\ref{fig:curve_compare} shows a comparison between the \textit{g}-band distribution of elliptical galaxies in this paper (as Fig.~\ref{fig:msrg}) and the sample of elliptical galaxies from \cite{Graham2006}. The red points show the distribution of our \textit{g}-band data and the red and green lines are the corresponding fits to Eqs.~\ref{equ:rm1} and \ref{equ:rm2} respectively.
The black triangles show the elliptical galaxies from \cite{Graham2006} and the curved purple line shows the expected relation presented in the same paper. The curved line is derived from considerations of the M$_{\rm gal}-\langle\mu\rangle _e$ relation \citep{Graham2003} and the luminosity relation, L$_{\rm gal}$ = 10$^{-M_{\rm gal}/2.5}$ = $2\left(\pi R_e^2 \langle I\rangle _e\right)$.

The sample of elliptical galaxies in \cite{Graham2006} is presented in B-band magnitudes, which have to be converted to stellar masses. To calculate the stellar mass from the given absolute B-band magnitudes we first convert from B-band to \textit{g}-band absolute magnitudes. According to Eq.~A5 in \cite{Cross2004} we have $B=\textit{g}+0.39(g-r)+0.21$ and we adopt the mean colour of our elliptical population of $(g-r)=0.71$.
The mass ($\mathcal{M}$) is simply given by
\begin{equation}
\mathcal{M}=\frac{\mathcal{M}}{L} 10^{0.4(M_{\sun}-M_{\rm gal})},
\end{equation}
where, instead of a constant mass-to-light ratio ($\frac{\mathcal{M}}{L} $), we use the mass and luminosities of our elliptical galaxies to establish the change of $\frac{\mathcal{M}}{L}$ with the absolute \textit{g}-band magnitude:
\begin{equation}
\left(\frac{\mathcal{M}}{L}\right)_g =10^{ -0.047 M_g - 0.608}.
\end{equation}
 
 It is obvious from Fig.~\ref{fig:curve_compare} that a curved relation is preferred when considering all elliptical galaxies from dwarf to giant ellipticals. Yet it is also clear that with the data available in our sample we do not observe enough low-mass $\left({\rm i.e.}~\mathcal{M}_*<\mathcal{M}_{lim}\right)$ galaxies to robustly constrain this curved relation.  Considering this we stress again that even though we recommend using the linear \msr relation fits to our data, these should only be compared to other data available in a similar mass range $\left({\rm i.e.}~10^9\mathcal{M}_{\sun}<\mathcal{M}_*<10^{11}\mathcal{M}_{\sun}\right)$. At lower masses a curved relation is preferred, but we caution that the curved fits provided in this paper are not well constrained for very low mass elliptical/ early-type galaxies.

\newpage
\section{Wavelength dependence of galaxy sizes}
\label{wavedepend}

\begin{figure*}
\includegraphics[width=1\textwidth]{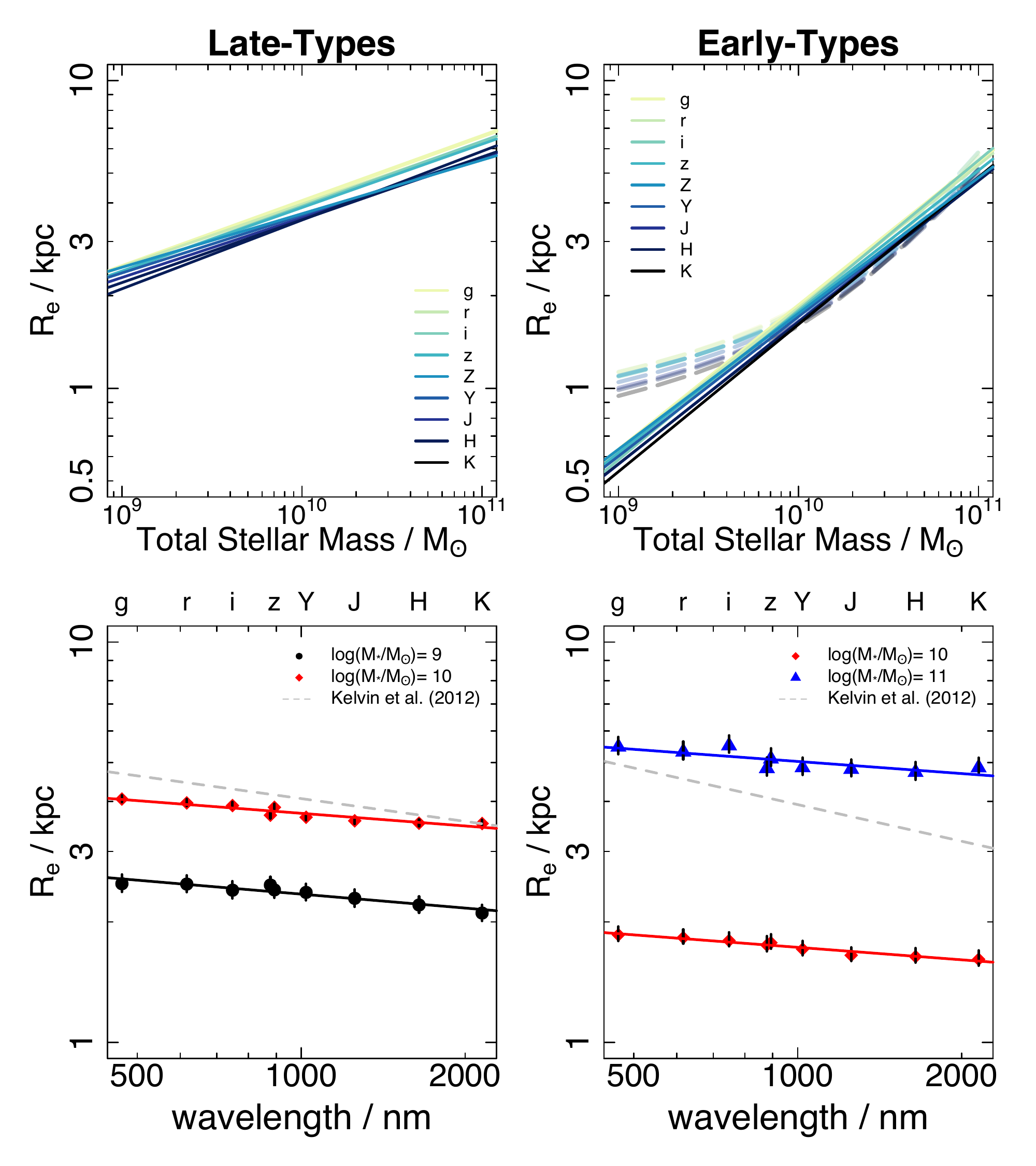}
\caption{The top panel shows the visually classified morphological late- and early-type \textit{grizZYJHK$_s$} \msr relations, in the left and right hand side plots respectively. We use the \msr relation fits to Eq.~\ref{equ:rm1} in our size-wavelength considerations, but the \msr relation fits to Eq.~\ref{equ:rm2} are shown for comparison for early-types galaxies in the top right plot (dashed, lighter coloured lines).
The bottom panel shows the corresponding size-wavelength variation with sizes calculated for different masses.
 We also show the best fit linear relation of the size-wavelength variation for each mass (fit parameters can be found in Table \ref{table:sizevar}). 
 In addition, we show the relations found by \citet{Kelvin2012} as the grey dashed line which was obtained over the entire mass range sampled in their paper. }
\label{fig:wave}
\end{figure*}

We investigate the wavelength dependence of galaxy sizes using the results of the \msr relation fits to our visually classified early-/ late-type sample.  
The top panel of Fig.~\ref{fig:wave} plots the \msr relation fits to Eq.~\ref{equ:rm1} in the \textit{grizZYJHK$_s$} bands for our late-types on the left and early-types on the right (we show the fits to Eq.~\ref{equ:rm2} as the dashed and lighter coloured lines, to illustrate the effect of the turn-off of the early-type \msr relation at lower masses).
It is clear that the early-type relation is steeper, and typically has smaller sizes, than the late-type relation with both \msr relations approaching similar sizes at $\mathcal{M}_*=10^{11}\mathcal{M}_{\sun}$. 
For both the late-type and early-type \msr relation we see a smooth progression of the expected size with wavelength from \textit{g}-band to  K$_s$-band.

However, it is also apparent that the \msr relations are not parallel and the offset is a function of stellar mass.
The bottom left panel of Fig.~\ref{fig:wave} plots the size change with wavelength for the late-types for two different masses, $\mathcal{M}_*=10^{9}\mathcal{M}_{\sun}$ and $10^{10}\mathcal{M}_{\sun}$. 
We did not investigate the size-wavelength trend at $\mathcal{M}_*=10^{11}\mathcal{M}_{\sun}$ since we can not constrain the \msr relation well due to small number statistics. 

Overall we observe a reduction in size (\textit{g} to K$_s$ band) for the late-types of 16\%, and 13\% at $\mathcal{M}_*=10^{9}\mathcal{M}_{\sun}$ and $10^{10}\mathcal{M}_{\sun}$ respectively. This is less than the size variation observed by \cite{Kelvin2012} and \cite{Vulcani2014}.
The best fit linear relations describing the size change in kpc with wavelength are shown in Table \ref{table:sizevar}. We have established a best fit linear relation for all masses probed and also present the relation found by \cite{Kelvin2012} for comparison. Please note that we did not correct our wavelengths to the restframe due to the limited redshift range sampled.\\

The bottom right panel of Fig.~\ref{fig:wave} plots the size change with wavelength for the early-types for two different masses, $\mathcal{M}_*=10^{10}\mathcal{M}_{\sun}$ and $10^{11}\mathcal{M}_{\sun}$. We did not investigate the expected size variation around $\mathcal{M}_*=10^{9}\mathcal{M}_{\sun}$ since our sample does not have a sufficient number of galaxies at low masses and hence the \msr relation is not well constrained.
For the early-types we observe a size reduction from \textit{g} to K$_s$ band of 13\% and 11\% at $\mathcal{M}_*=10^{10}\mathcal{M}_{\sun}$ and $10^{11}\mathcal{M}_{\sun}$. This is significantly less than the change reported in \cite{Kelvin2012} who found a size reduction of 38\% for their full early-type sample. \\

\begin{table}
\centering
\begin{tabular}{|l|l}\\
	\hline
	\multicolumn{2}{l}{a) Late-type size-wavelength variation}\\
Case 	&  relation \\
	\hline \hline
	\\
		10$^9{ }$M$_{\sun}$		&   $\log_{10}(R_e)=  -0.116$  $\log_{10}(\lambda) + 0.717$ \\
		10$^{10}$M$_{\sun}$		&	 $\log_{10}(R_e)=  -0.105$  $ \log_{10}(\lambda) + 0.887$	 \\
		\cite{Kelvin2012}				&	 $\log_{10}(R_e)=  -0.189$  $\log_{10}(\lambda_{rest}) + 1.176$\\
\hline
\\
\end{tabular}
\begin{tabular}{|l|l}\\
	\hline
	\multicolumn{2}{l}{b) Early-type size-wavelength variation} \\
Case 	& relation \\ 
	\hline \hline
	\\
		10$^{10}$M$_{\sun}$		& 		 $\log_{10}(R_e)=  -0.104$  $ \log_{10}( \lambda) + 0.548$ \\
		10$^{11}$M$_{\sun}$		&		 $\log_{10}(R_e)=  -0.101$  $ \log_{10}(\lambda) + 1.004$ 	\\	
		\cite{Kelvin2012}				&	  	 $\log_{10}(R_e)=  -0.304$  $ \log_{10}(\lambda_{rest})+ 1.506$\\
			\hline
\end{tabular}
\caption{The size-wavelength variations of late and early-types for different masses.}
\label{table:sizevar}
\end{table}

We believe that this reduction in observed size variation, both for early- and late-types, is due to the switch from the shallower UKIDSS \textit{YJHK} imaging data to the deeper VIKING \textit{YJHK$_s$} imaging data.
The spheroid population typically has high S\'ersic index values (i.e. n$\sim$4) which means they have very extended lower surface brightness wings which can lead to an overestimation of the local sky level. However, with the improvement of the imaging data switching from UKIDSS to VIKING these galaxy wings become detectable above the noise level and we recover larger radii during the light profile fitting.
Galaxy discs are less affected by this since their low S\'ersic index (n$\sim$1) means they do not have low surface brightness wings which contain a significant flux contribution. \\

We attribute the observed size variation in the late-type galaxies to dust attenuation which would preferentially obscure the central regions of galaxies and thus cause an artificial shift to higher effective half-light radii in the shorter optical bands. As such, this effect should be more prominent in disc galaxies which are dustier than spheroid galaxies. 
In fact the observed size change is in agreement with expected values, see e.g. \cite{Pastrav2013a} who predict an effect of $\sim$15\% on the sizes of discs due to dust attenuation. 
However, the observed size drop of $\sim$13\% in our spheroid sample, which typically have no dust associated with them, suggests that other effects also influence the observed size variation of galaxies, such as stellar population or metallicity gradients and the two component nature of galaxies (see \citealt{Vulcani2014}, who have also noted this), i.e. generic inside-out formation histories with discs continually growing through gas infall and spheroids accreting in minor merger events.\\

It is also interesting to note that we see a slight decrease of the size-wavelength dependence of galaxies with increasing mass across the early- and late-types. We are not certain if this trend is real (we only sample a small number of masses) nor do we fully understand the cause of this trend, if it is indeed significant. 
However, it would generally be consistent with downsizing (i.e. massive systems form faster, \citealt{Thomas2005}). In this context the massive galaxies are likely to be the oldest in our sample and hence would have had more time to re-distribute their stellar populations (in part aided by major mergers, see e.g. \citealt{Conselice2014}) so that we see less stellar population (or colour) gradients and hence their R$_e$ and S\'ersic index should change less with wavelength. 
In contrast for the less massive, and probably younger, galaxies we potentially see the traces of their accretion history (including minor mergers), where we have an older more centrally concentrated stellar populations and a younger more wide spread stellar population. This would be in accordance with the inside-out growth scenario for galaxies (e.g. \citealt{Hopkins2009}).

\section{Discussion and Summary}

We use a sample of GAMA galaxies with redshifts between
0.01$\leq$z$\leq$0.1 and magnitudes of $r<$19.8 mag to study the \msr relation in the \textit{ugrizZYJHK$_s$} bands. To establish a comprehensive set of z=0 \msr relations we first set up a common sample of 8399 (6154) galaxies in the $g-K_s$ ($u-K_s$) bands with high quality galaxy profiles in all bands. We also carefully consider our selection boundaries and find that our data lies within the observable window, allowing for an unbiased fit to the \msr relation. Furthermore we split the sample into early and late-type using several common separators:
\begin{enumerate}
\item the S\'ersic index, 
\item the dust corrected restframe (\textit{u-r})$_{stars}$ colour, 
\item the dust corrected restframe (\textit{g-i})$_{stars}$ colour, 
\item a combined (\textit{u-r})$_{stars}$ colour-S\'ersic  index cut and 
\item the visual morphology of galaxies.
\end{enumerate} 

The resulting early- and late-type populations are fit with two functions, a single power law and a two-component power law.
For the late-type samples, the two-component power law shows some variation with the `transition mass' M$_0$ changing significantly for different chosen separators. This is not surprising since we leave M$_0$ as a free parameter in contrast to the S03 fits where M$_0$ was set at the point at which the dispersion of their data changes and moves from `high' to `low' mass galaxies.
As such M$_0$ is only an artificial `transition mass' in our fits and no real physical meaning can be assigned.
In addition, we find that most parameters of Eq.~\ref{equ:rm2} change significantly with the different cuts whereas parameter b in Eq.~\ref{equ:rm1} stays remarkably constant and only the intercept changes with the chosen separator (indicating the biases introduced by the different separators).
Considering this we find that the single power law fit is sufficient to describe the data and recommend using it as the canonical reference in comparison with other data sets. 

For the early-types however we find that the two-component power law has more robust results than the single-component power law due to the flattening of the \msr distribution. We recommend that a move to a curved relation for the elliptical (early-types) galaxies is necessary (such as seen in \citealt{Graham2006}). However, if mostly high-mass elliptical galaxies are studied a single-component power law may be sufficient, but we caution that the slope for the single-component power law in Table \ref{table:rmfitsE} describes the overall sample and hence is too shallow to describe a sample of only high-mass galaxies adequately. For those cases when a linear comparison is needed we provide additional \msr relations for early-type galaxies with masses of  $\mathcal{M}_*>\mathcal{M}_{lim}=2\times10^{10}\mathcal{M}_{\sun}$ (fit to Eq.~\ref{equ:rm1}) in Appendix \ref{app:msr_additional}.\\

Table \ref{table:scatter} shows the percentage of galaxies that have been correctly classified as early- or late-type according to our visual classification and the overall likelihood that a galaxy is correctly identified as either early- or late-type. We have divided our sample into high and low mass galaxies using the mass limit established for a volume limited sample (i.e. low mass galaxies have masses of $\mathcal{M}_*<\mathcal{M}_{lim}=10^{9.4}\mathcal{M}_{\sun}$). This allows us to better quantify which separator performs best and at which mass range the most problems are encountered.

On the basis that we want the most reliable selection for a sample of morphological late type galaxies we find that the $(g-i)_{stars}$ colour cut performs the best at higher masses (Table \ref{table:scatter}b, c) and the S\'ersic index performs best at low masses (Table \ref{table:scatter}a). 
The most reliable early-type selection is given by the rolling cut for higher masses (Table \ref{table:scatter}b, c) and the $(g-i)_{stars}$ colour at low at masses (Table \ref{table:scatter} a). 
The rolling cut failed at the low masses due to low number statistics. Since, by definition, the rolling cut maximizes the bijective probability of the galaxies being correctly identified as early- or late-type (in terms of morphology) it is biased towards the higher mass galaxies where most ellipticals can be correctly identified. 
Hence out of the 35 low-mass galaxies identified as early-type using the rolling cut none of them are found to be elliptical galaxies.\\
We find that a S\'ersic-index selection is the least reliable selection that we have considered for discriminating between morphological early- and late-type galaxies. 
The inspection of the low mass ($\mathcal{M}_*<\mathcal{M}_{lim}=10^{9.4}\mathcal{M}_{\sun}$) cross-scatter seen in the early-type sample using the S\'ersic index cut shows that these galaxies are predominantly blue in colour (i.e. $(u-r)_{stars}<1.5$), have a median size R$_e\sim1.2$kpc and have comparably low S\'ersic indices (that is 53$\%$ have a S\'ersic index n$<$3 as opposed to only 23$\%$ for the entire early-type sample) hence most of this population was likely missed in the S03 analysis.

Consequently the low mass galaxies which are moved into our early-type sample by the S\'ersic index cut cause a flattening in the \msr distribution. This flattening is not unlike the that seen for the elliptical galaxies but should not be confused with it since in the low mass cross-scatter is predominantly made up of morphological late-type galaxies. The flattening of the \msr relation fit could become even more significant when using other (less robust) fitting routines or further expanding the low mass end of the data set. We advise caution when considering the S\'ersic index to split a data set into early- and late types especially if the early-type galaxies are of particular interest and low mass galaxies are included.\\

 Even using our simple morphological classification of elliptical and not-elliptical to distinguish the early and late-type galaxies, we can see a correlation with colour and S\'ersic index, but they are by no means synonymous with the morphological classification.
Using generic\slash rigid separators to divide the galaxy
population into early- and late types should be used with caution and most importantly wherever possible the same separation schemes should be compared.  If morphological
information is unavailable both a division by dust corrected colour or a combined S\'ersic and colour division are good
alternatives to separate the early- and late-type galaxies. We find that our $(g-i)_{stars}$ colour performs slightly better than our $(u-r)_{stars}$ colour. However, this is likely an effect of the poorer imaging quality in the \textit{u}-band which translates to a slightly less reliable $(u-r)_{stars}$ colour.
Overall, the S\'ersic index is the least desirable separator, especially if the sample extends to lower masses (see Fig.~\ref{fig:select}).\\

\begin{table}
\centering
\begin{tabular}{|l|l|l|l|l|l}\\
	\hline
	a) $\mathcal{M}_*<\mathcal{M}_{lim}$\\
Case 			& late-type & $\times$ & early-type & = & bijective\\ 
Sample size & 87.8\% 		&&	2.6\% \\
	\hline \hline
	\\
		S\'ersic index				& 		\textbf{0.896} 		&&		0.139 					&&	0.125 \\
		(\textit{u-r})$_{stars}$	&		0.882 					&&		0.189 					&&	0.167 \\
		(\textit{g-i})$_{stars}$	&		0.883					&&		\textbf{0.242} 		&&	\textbf{0.214} \\
		rolling cut					&		0.878 					&&		{0} 				&&	{0}\\
\end{tabular}
\begin{tabular}{|l|l|l|l|l|l}\\
	\hline
	b) $\mathcal{M}_*>\mathcal{M}_{lim}$ \\
Case 	& late-type & $\times$ & early-type & = & bijective\\ 
Sample size & 70.5\% 		&&	27.9\% \\
	\hline \hline
	\\
		S\'ersic index				& 		{0.874} &&		0.653 &&	0.57 \\
		(\textit{u-r})$_{stars}$	&		0.879 &&		0.623 &&	0.548 \\
		(\textit{g-i})$_{stars}$	&		\textbf{0.881} &&		0.614 &&	0.544 \\
		rolling cut					&		0.835 &&		\textbf{0.723} &&	\textbf{0.604} \\
\end{tabular}
\begin{tabular}{|l|l|l|l|l|l}\\
	\hline
	c) entire sample\\
Case 	& late-type & $\times$ & early-type & = &  bijective \\ 
Sample size & 73.2\% 		&&	23.9\% \\
	\hline \hline
	\\
		S\'ersic index				& 		{0.879} &&		0.636 &&	0.559 \\
		(\textit{u-r})$_{stars}$	&		0.88 &&		0.616 &&	0.542 \\
		(\textit{g-i})$_{stars}$	&		\textbf{0.881} &&		0.613 &&	0.54 \\
		rolling cut					&		0.844 &&		\textbf{0.722} &&	\textbf{0.61}\\
\end{tabular}
\caption{The table shows the fraction of late- and early-type galaxies which are classified as not-elliptical or elliptical for the four rigid population cuts. In addition we calculate the (bijective) probability of any galaxy in the sample having been correctly associated with the morphological early- and late-type classifications.
From top to bottom we show this for galaxies a) below the mass limit; b) galaxies above the mass limit and c) the entire sample. In each case we also show the percentage of the entire sample that were visually classified as either elliptical (early-type) or not-elliptical (late-type). The sample sizes do not add up to 100\% and the missing fraction is represented by the LBS.}
\label{table:scatter}
\end{table}

In addition to the various population separators we have analysed the \msr relation in 10 imaging bands, \textit{ugrizZYJHK$_s$}. Fitting in each band is done for all five population separators using the fitting routines as described for the \textit{r}-band data in Sec.~\ref{sec:MSR}.
This is important for various reasons such as the change in population make-up when using non-morphological early-\slash{}late-type cuts. The most noticeable effect is the observed change in galaxy size with wavelength \citep{LaBarbera2010,Kelvin2012,Haussler2013,Vulcani2014,Wel2014}, which could be caused by dust attenuation and/or the inside out growth of galaxies which causes different stellar populations to be observed at different wavelengths and hence is an effect of both a change in colour as well as S\'ersic index.
This effect will also be important when comparing to high redshift data due to the shift in restframe wavelength. It is therefore imperative to take the change in size, as well as the population make-up due to colour and S\'ersic index changes, into account when studying the growth of galaxies.\\

Finally, we have studied the size-wavelength dependence using the \msr relation fits to the  \textit{grizZYJHK$_s$} for early- and late-types as classified by their visual morphologies.
We confirm the presence of a size-wavelength dependence for both early- and late-type galaxies.
However, we find that the previously reported size drop of 38\% for early-types \citep{Kelvin2012} has likely been an overestimation which can be attributed to the limiting near-IR imaging data quality.  
In our analysis we have used VIKING \textit{ZYJHK$_s$} instead of UKIDSS \textit{YJHK} band imaging data and find that late-type galaxies experience an average size drop of $\sim$14\% and early-type galaxies a size drop of $\sim$12\%, much less than previously reported. \\
It is also interesting to note that the observed change in galaxy size with wavelength might depend on the mass-range probed. However, this trend needs further investigation and might actually depend on the (imaging) quality of the data. \\

 In this paper we have presented the \msr relation for local galaxies in 10 imaging bands, \textit{ugrizZYJHK$_s$}, using five different early-\slash{}late-type separators to split the galaxy population. This extensive collection of various \msr relations should allow for the convenient comparison of our local \msr relation with other local relations as well as high redshift relations using the same restframe wavelength population separation criteria.

 In future work we will expand our analysis to look in more detail at disks, spheroids and galaxy components.
 The study of the \msr relation by galaxy type and component will lay the foundations for more thorough comparisons with intermediate to high redshift data. For example, it has been put forward that compact high redshift galaxies are actually the cores of modern day galaxies (see e.g. \citealt{Driver2013,Dullo2013}). To confirm this, it is necessary to establish a robust \msr relations of local galaxy components for comparison.
Additionally, we can further test evolutionary models by studying the connection between angular momentum and galaxy size, more specifically we will study the mass-spin-morphology relation using the disc \msr relation of galaxies.

\section{Acknowledgements}
RL would like to acknowledge funding from the International Centre for
Radio Astronomy Research and the University of Western Australia.  
 GAMA is a joint European-Australasian project based around a spectroscopic campaign using the Anglo-Australian Telescope. 
 The GAMA input catalogue is based on data taken from the Sloan Digital Sky Survey and the UKIRT Infrared Deep Sky Survey. 
 Complementary imaging of the GAMA regions is being obtained by a number of independent survey programs including GALEX MIS, VST KiDS, VISTA VIKING, WISE, Herschel-ATLAS, GMRT and ASKAP providing UV to radio coverage. 
 The VISTA VIKING data used int his paper is based on observations made with ESO Telescopes at the La Silla Paranal Observatory under programme ID 179.A-2004.
GAMA is funded by the STFC (UK), the ARC (Australia), the AAO, and the participating institutions. The GAMA website is http://www.gama-survey.org/.

\clearpage

\footnotesize
\bibliographystyle{mn2e}
\setlength{\bibhang}{2.0em}
\setlength{\labelwidth}{0.0em}
\bibliography{GAMA_MSR_bib}
\normalsize

\appendix

\section{The SDSS \textit{ugiz} and VIKING \textit{ZYJHK$_s$} \msr relations}
\label{app:msr_allband}

We have calculated the \msr relation in 10 available imaging bands. 
The \textit{r}-band relations for a S\'ersic index cut, two colour cuts, a combined S\'ersic index and colour cut as well as a morphologically classified sample were shown in the main part of the paper. 
Here we present the \msr relations for the additional 9 bands \textit{ugizZYJHK$_s$}. \\
The data was analysed in the same way as outlined for the \textit{r}-band data. 
We exclude outliers and bad fits in each band individually in addition we also remove galaxies with unrealistic fitting parameters which leads to varying sample sizes. However, after implementing the staggered volume limited sample in each band the final sample sizes, with the exception of the \textit{u}-band, are comparable with each other. The number of `good-fit' galaxies and the staggered volume limited sample size for each band can be found in table \ref{table:samplesize} and Tables \ref{table:rmfitsL} and \ref{table:rmfitsE} show the resulting fitting parameters to  Eqs. \ref{equ:rm1} and \ref{equ:rm2} for the late and early-types respectively.
The following 9 plots are equivalent to the \textit{r}-band plot presented in the main part of the paper and show the \msr relations for \textit{ugizZYJHK$_s$} late types (left hand panels, blue) and early-types (right hand panels, red) divided from top to bottom panel by:
\begin{enumerate}
\item the S\'ersic index, 
\item the restframe (\textit{u-r}) colour, 
\item the restframe (\textit{g-i}) colour,
\item a combined (\textit{u-r})$_{stars}$ colour-S\'ersic index cut and 
\item the visual galaxy morphology.
\end{enumerate} 

\begin{figure*}
	\centering
	\includegraphics[width=0.9\textwidth]{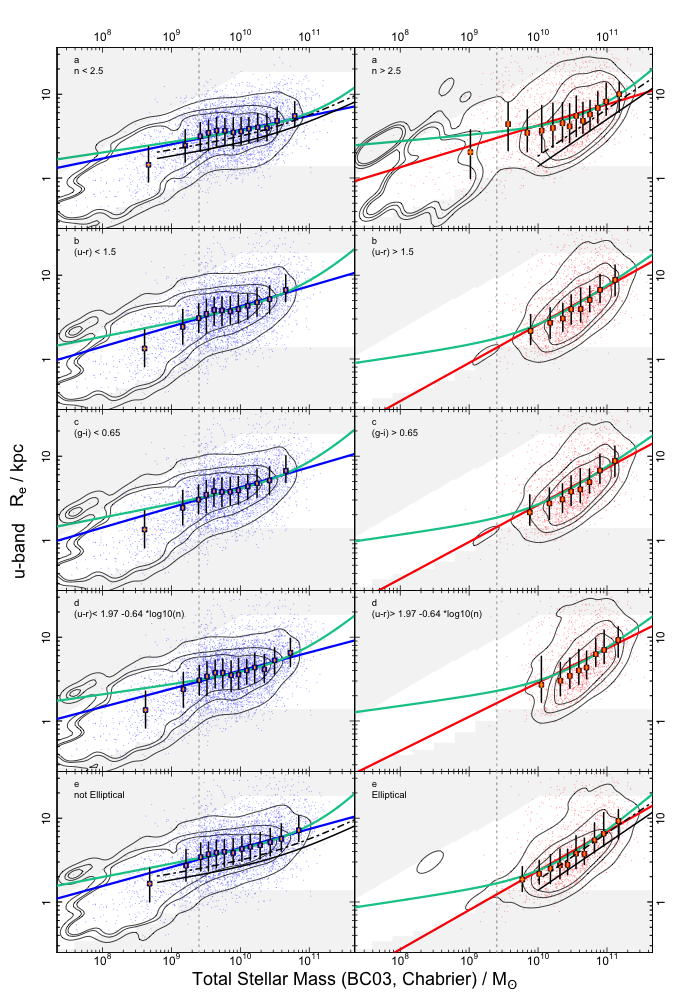}
	\caption{The \textit{u}-band \msr relation for late and early-types - left and right hand side respectively. We are using a) the S\'ersic index, b) the \textit{(u-r)}$_{stars}$ colour, c) the  \textit{(g-i)}$_{stars}$ colour, d) a combination of S\'ersic index and \textit{(u-r)}$_{stars}$ colour and e) the visual morphology to divide the populations as described in the paper. Fitting parameters can be found in Tables \ref{table:rmfitsL} and \ref{table:rmfitsE}.}
	\label{fig:msru}
\end{figure*}

\begin{figure*}
	\centering
	\includegraphics[width=0.9\textwidth]{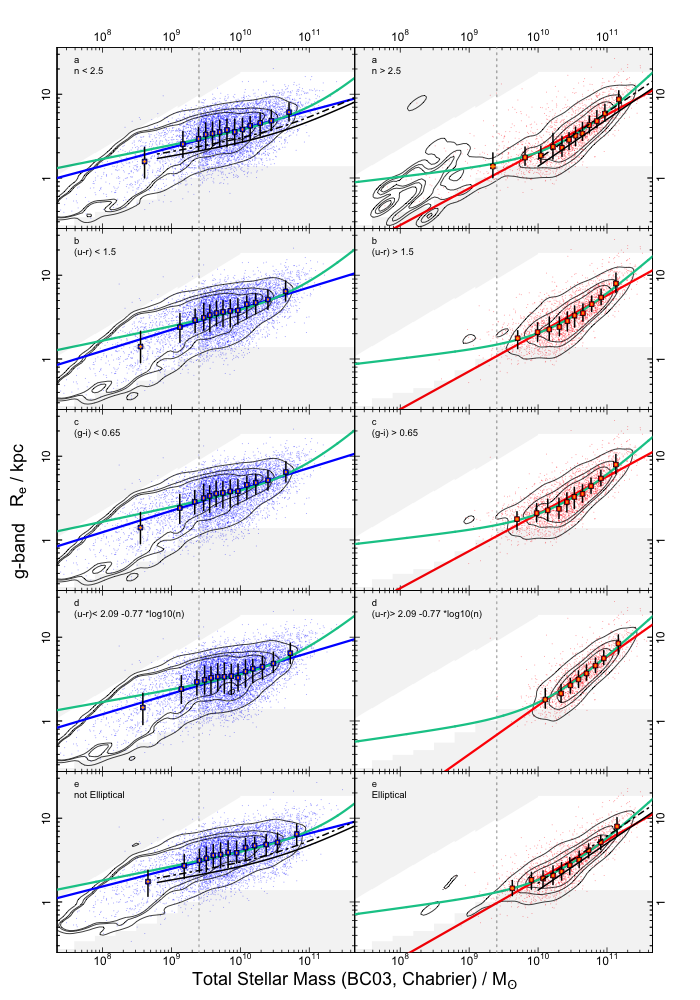}
	\caption{The \msr relation for the \textit{g}-band with the panels and fits as in Fig.~\ref{fig:msru}.}
	\label{fig:msrg}
\end{figure*}

\begin{figure*}
	\centering
	\includegraphics[width=0.9\textwidth]{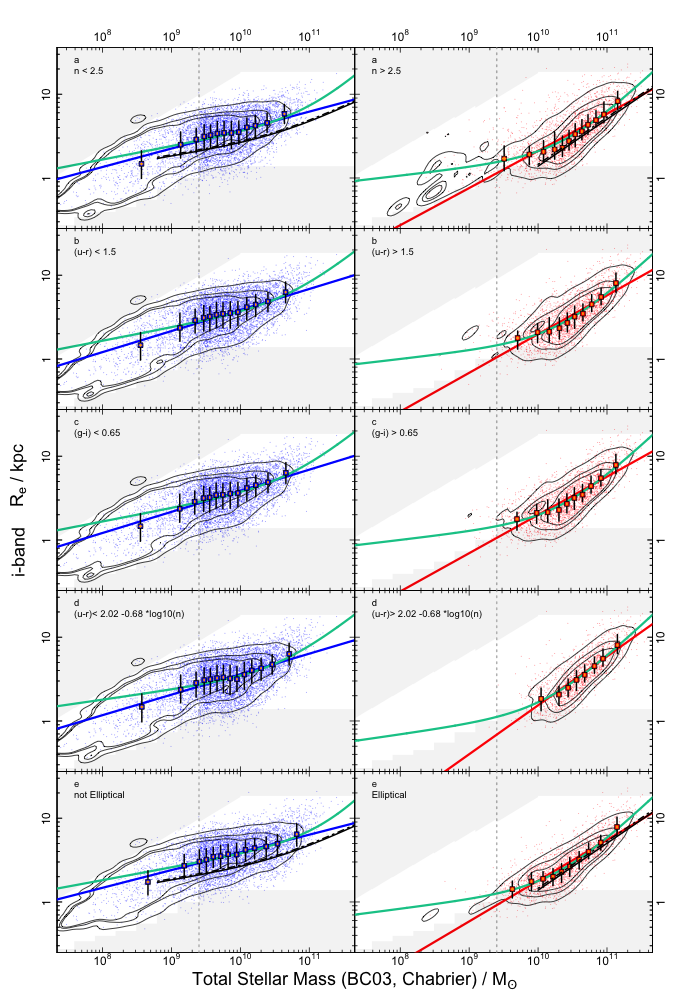}
	\caption{The \msr relation for the \textit{i}-band with the panels and fits as in Fig.~\ref{fig:msru}.}
	\label{fig:msri}
\end{figure*}

\begin{figure*}
	\centering
	\includegraphics[width=0.9\textwidth]{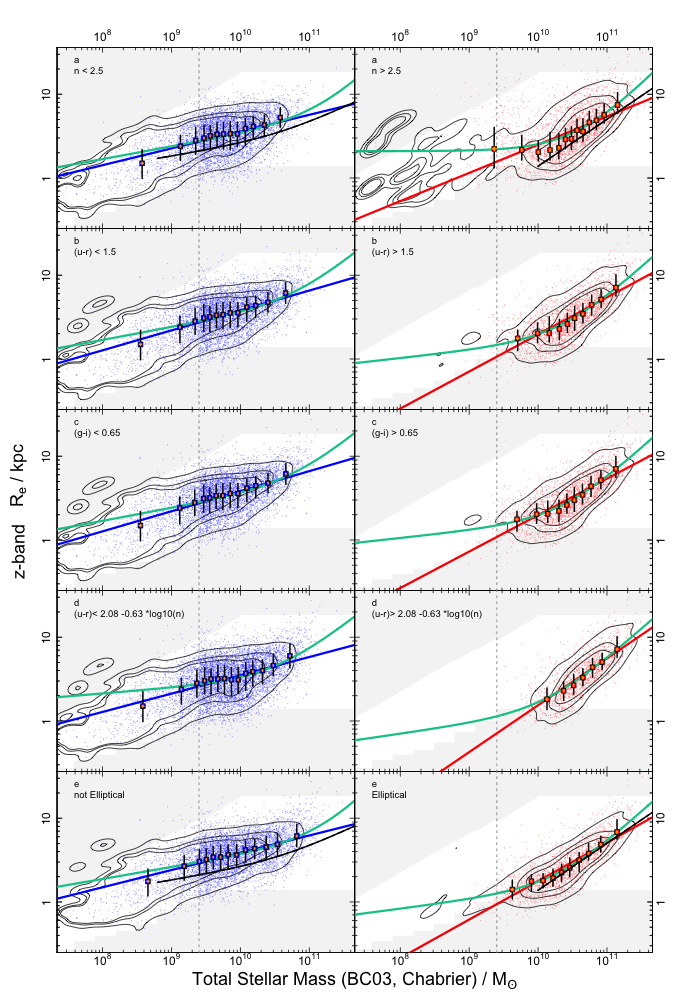}
	\caption{The \msr relation for the \textit{z}-band with the panels and fits as in Fig.~\ref{fig:msru}.}
	\label{fig:msrz}
\end{figure*}

\begin{figure*}
	\centering
	\includegraphics[width=0.9\textwidth]{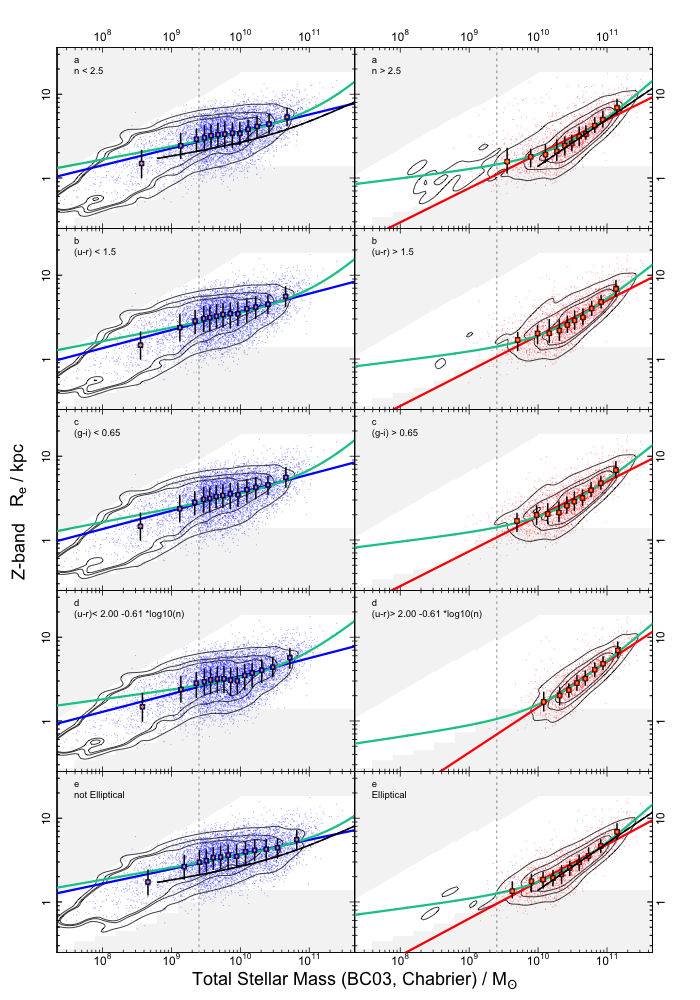}
	\caption{The \msr relation for the VIKING \textit{Z}-band with the panels and fits as in Fig.~\ref{fig:msru}.}
	\label{fig:msrzv}
\end{figure*}

\begin{figure*}
	\centering
	\includegraphics[width=0.9\textwidth]{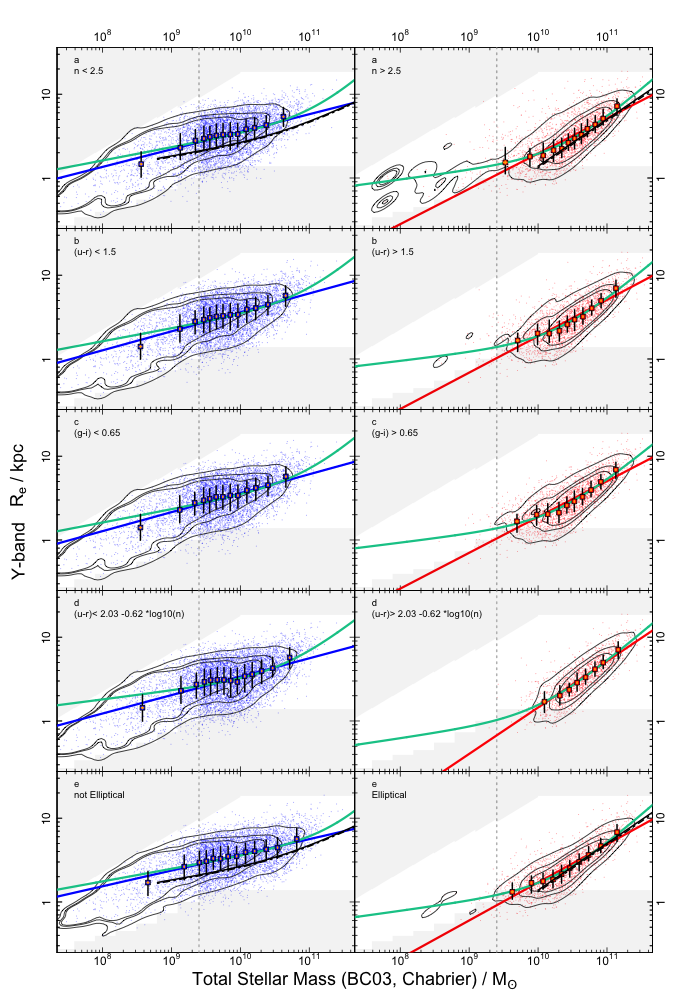}
	\caption{The \msr relation for the \textit{Y}-band with the panels and fits as in Fig.~\ref{fig:msru}.}
	\label{fig:msry}
\end{figure*}

\begin{figure*}
	\centering
	\includegraphics[width=0.9\textwidth]{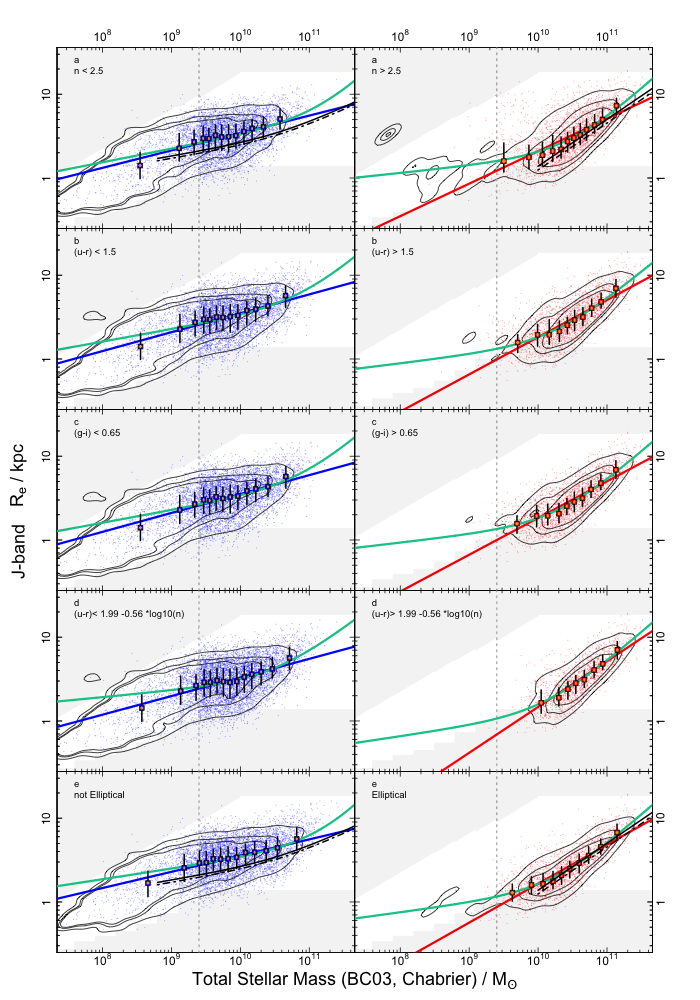}
	\caption{The \msr relation for the \textit{J}-band with the panels and fits as in Fig.~\ref{fig:msru}.}
	\label{fig:msrj}
\end{figure*}

\begin{figure*}
	\centering
	\includegraphics[width=0.9\textwidth]{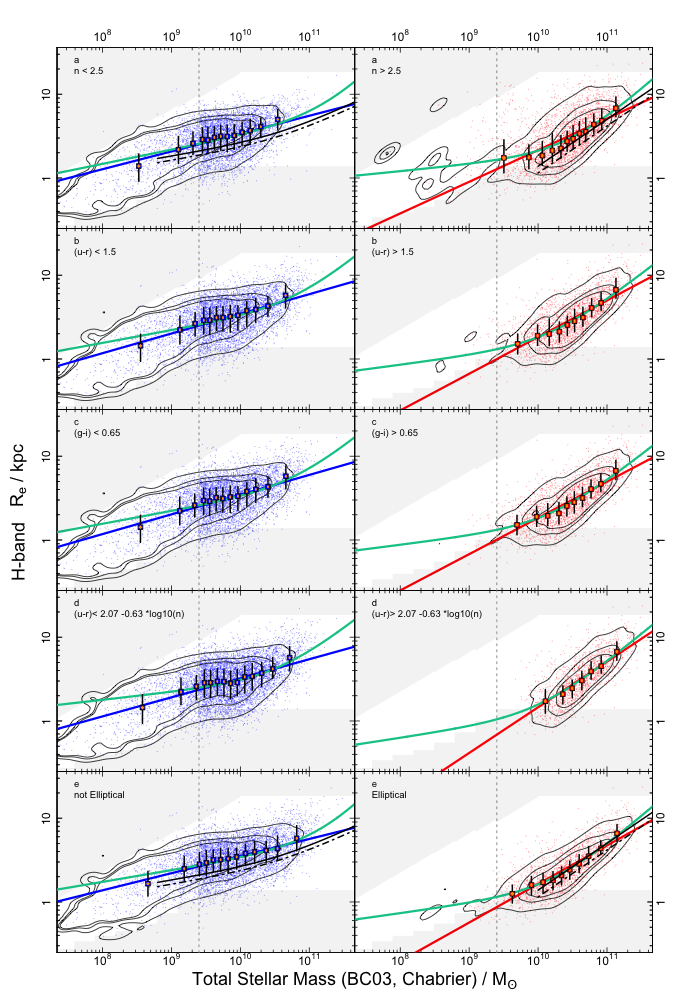}
	\caption{The \msr relation for the \textit{H}-band with the panels and fits as in Fig.~\ref{fig:msru}.}
	\label{fig:msrh}
\end{figure*}

\begin{figure*}
	\centering
	\includegraphics[width=0.9\textwidth]{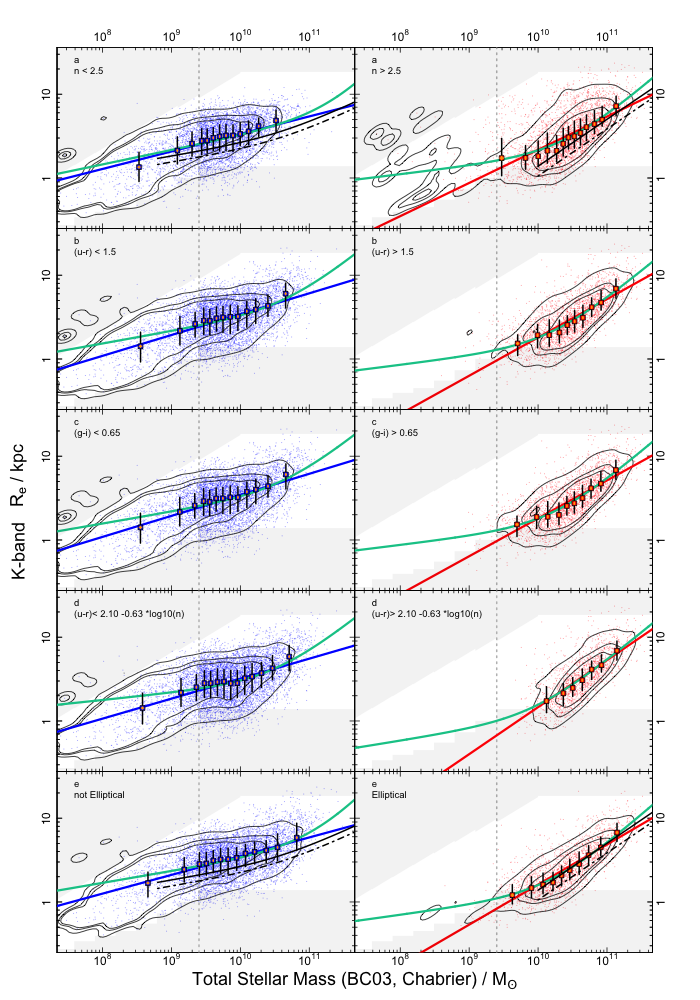}
	\caption{The \msr relation for the \textit{K}-band with the panels and fits as in Fig.~\ref{fig:msru}.}
	\label{fig:msrk}
\end{figure*}

\section{Additional \msr relations}
\label{app:msr_additional}

In high redshift studies it becomes more difficult to divide the sampled galaxies into the conventional early- and late-types.
To allow a more direct comparison to high redshift data we have fit the local \msr relation using Eq.~\ref{equ:rm1} to the entire sample without any early-/ late-type division. The results for all 10 imaging bands can be found in Table \ref{table:msrall}.\\

In addition, in many cases (especially at higher redshifts) only high-mass data is available to establish a \msr relation. In the case of the early-type galaxies this  exclusion of low-mass data can lead to a significant change in the slope of the \msr relation (since a single-power law is sufficient to describe the data). 
To allow for easier comparison to these high-mass (and/ or high-z) data we have analysed the \msr relation of our high-mass early-types. For this we fit the \msr relation to elliptical galaxies with masses $\mathcal{M}_*>2\times10^{10}\mathcal{M}_{\sun}$. This is the average transition mass M$_0$ according to our morphology cut (\textit{g-K$_s$} band, see Table \ref{table:rmfitsE}) and also agrees with the limit imposed by \citealt{Wel2014} to avoid the flattening of the early-type relation. 

We also set up a high-mass sample for the 'not-elliptical' galaxies. For this we set the lower mass limit to the mass limit of our colour unbiased volume limited sample, $\mathcal{M}_*>2.5\times10^{9}\mathcal{M}_{\sun}$. Not unsurprisingly the results of the \msr relation fit remain mostly unchanged. This is in good agreement with our previous observation that the late-type \msr relation is fairly robust to changes in the population set up.
The resulting fitting parameters can be found in Table \ref{table:msr_highm}.

\begin{table}
\centering
\begin{tabular}{|l|l|l|l}\\
	\hline
Band 			&& a  (10$^{-2}$) & b \\ 
	\hline \hline
	\\
u   && 1.79 $\pm$ 0.17  & 0.23 $\pm$ 0.02\\
g   && 3.43 $\pm$ 0.35  & 0.20 $\pm$ 0.01\\
r   && 4.04 $\pm$ 0.42  & 0.19 $\pm$ 0.01\\
i   && 2.86 $\pm$ 0.27  & 0.21 $\pm$ 0.01\\
z   && 4.02 $\pm$ 0.42  & 0.19 $\pm$ 0.01\\
Z   && 5.99 $\pm$ 0.77  & 0.17 $\pm$ 0.01\\
Y   && 4.73 $\pm$ 0.51  & 0.18 $\pm$ 0.01\\
J   && 4.08 $\pm$ 0.44  & 0.19 $\pm$ 0.01\\
H   && 3.59 $\pm$ 0.37  & 0.19 $\pm$ 0.01\\
K   && 2.43 $\pm$ 0.22  & 0.21 $\pm$ 0.01\\
\hline
\end{tabular}
\caption{\msr relation fitting parameters to Eq.{}\ref{equ:rm1} for the entire sample without any early-/ late-type division.}
\label{table:msrall}
\end{table}
   
  \begin{table}
\centering
\begin{tabular}{|l|l|l|l}\\
	\hline
  \multicolumn{2}{l}{Late-types} \\
Band					&& a  (10$^{-2}$) & b \\ 
	\hline \hline
	\\
g  && 3.32 $\pm$ 0.44  & 0.21 $\pm$ 0.02  \\
r  && 4.02 $\pm$ 0.57  & 0.20 $\pm$ 0.02 \\
i  && 3.04 $\pm$ 0.40  & 0.21 $\pm$ 0.02 \\
z  && 3.45 $\pm$ 0.47  & 0.21 $\pm$ 0.02 \\
Z  && 7.27 $\pm$ 1.25  & 0.17 $\pm$ 0.02  \\
Y  && 4.98 $\pm$ 0.75  & 0.19 $\pm$ 0.02  \\
J   && 4.27 $\pm$ 0.61  & 0.19 $\pm$ 0.02  \\
H  && 3.23 $\pm$ 0.45  & 0.20 $\pm$ 0.02  \\
K  && 2.07 $\pm$ 0.25  & 0.22 $\pm$ 0.02  \\
\hline
\end{tabular}
\\
\begin{tabular}{|l|l|l|l}\\
\hline 
  \multicolumn{2}{l}{Early-types} \\
		Band			&& a  (10$^{-6}$) & b \\ 
					\hline \hline
g  &&  0.63 $\pm$ 0.05  & 0.63 $\pm$ 0.03\\
r  &&  0.79 $\pm$ 0.06  & 0.62 $\pm$ 0.03\\
i  &&  0.35 $\pm$ 0.03  & 0.66 $\pm$ 0.03\\
z  &&0.85 $\pm$ 0.07  & 0.62 $\pm$ 0.03\\
Z  &&  1.36 $\pm$ 0.12  & 0.60 $\pm$ 0.03\\
Y  &&  1.25 $\pm$ 0.11  & 0.60 $\pm$ 0.03\\
J   &&  0.96 $\pm$ 0.08  & 0.61 $\pm$ 0.03\\
H  &&  1.46 $\pm$ 0.13  & 0.59 $\pm$ 0.03\\
K  &&  0.92 $\pm$ 0.08  & 0.61 $\pm$ 0.03\\
\hline
\end{tabular}
\caption{\msr relation fitting parameters to Eq.~\ref{equ:rm1} for high mass morphological late- and early- type galaxies with $\mathcal{M}_*>2.5\times10^{9}\mathcal{M}_{\sun}$ and $\mathcal{M}_*>2\times10^{10}\mathcal{M}_{\sun}$ respectively.}
\label{table:msr_highm}
\end{table} 

\label{lastpage}
\end{document}